\documentclass[useAMS,usenatbib]{mn2e}

\usepackage{natbib}
\usepackage{psfig}
\usepackage{amssymb}
\usepackage{longtable}
\usepackage[utf8]{inputenc}
\usepackage{supertabular}

\newcommand{\be}{\begin{equation}}
\newcommand{\en}{\end{equation}}

\def\mgii{Mg~{\sc ii}}

\def\ni2{Ni~{\sc ii}}

\def\civ{C~{\sc iv}}
\def\siiv{Si~{\sc iv}}

\def\si2{Si~{\sc ii}}

\def\h1{H~{\sc i}}

\def\kms{kms$^{-1}$}
\title[Transient \civ\ Broad Absorption Lines]{Transient \civ\ Broad Absorption Lines in radio detected QSOs}
\author[M. Vivek et al.]{M. Vivek$^{1}$\thanks{E-mail:vivekm@astro.utah.edu},R. Srianand$^{2}$ \& N. Gupta$^{2}$\\
$^{1}$ Department of Physics and Astronomy, University of Utah, Salt Lake City, UT 84112, USA\\
$^{2}$Inter University Centre for Astronomy and Astrophysics, Pune 410007, India}
\begin{document}
\date{Accepted . Received ; in original form }
\pagerange{\pageref{firstpage}--\pageref{lastpage}} \pubyear{2002}
\maketitle

\label{firstpage}
\begin{abstract}
 We study the transient (i.e. emerging or disappearing) \civ\ broad
absorption line (BAL) components in 50 radio detected QSOs using multi-epoch 
spectra available in Sloan Digital Sky Survey DR10. We report the detection
of 6 BALQSOs having at least one distinct transient \civ\ absorption component. Based
on the structure function analysis of optical light curves, we suggest
that the transient absorption is unlikely to be triggered by continuum 
variations. Transient absorption components usually have low \civ\ 
equivalent widths ($<$8 \AA), high ejection velocities ($>$ 10000 \kms) 
and typically occur over rest-frame timescales $>$ 800 days. 
The detection rate of transient \civ\ absorption 
seen in our sample is higher than that reported in the literature. 
Using a control sample of QSOs, we show that
this difference is most likely due to the longer monitoring time-scale of sources
in our sample  while the effect of small number statistics cannot be ignored. 
Thus, in order to establish the role played by radio jets in driving the BAL outflows, we need a 
larger sample of radio detected BALs monitored over more than 3 years in the 
QSO's rest frame. We also find that the transient phenomenon in radio detected and radio quiet 
BALs does not depend on any of the QSO properties i.e. the Eddington ratio, 
black hole mass, bolometric luminosity and optical-to-IR colours. 
All this suggests that transient BAL phenomenon is simply the extreme case of BAL variability. 
%
\end{abstract}


\section{Introduction}
Outflows in QSOs are detected as blue-shifted broad absorption line (BAL) 
troughs \citep[see][]{Weymann91} in the optical spectrum of  20-40\% of QSOs. 
This observed  BAL incidence is  interpreted as the covering factor of the 
BAL outflow in the orientation based models  and as the duration of  
BAL phase in a QSO's life in the evolutionary models.
Photoionization models with additional constraints on electron density allow
one to determine the location of the outflow and estimate the 
mass outflow rate. 
The models suggest that the outflows 
carry sufficient kinetic luminosity to provide the necessary Active Galactic Nuclei (AGN) feedback 
often invoked in theoretical and numerical simulations to explain the observed properties of 
galaxies \citep{moe09, dunn10, bautista10,borguet13}. Thus, understanding the
origin and physical conditions in BAL outflows is not only important for
understanding physics of AGNs but also for understanding feedback associated with the galaxy evolution \citep[][]{germain09,hopkins09}.

Radio jets and lobes are  manifestations of large scale outflows 
of relativistic plasma from central regions of the QSOs. It is well established that 
by studying radio emission at different frequencies and angular scales
one can statistically constrain the orientation and the age of radio sources. 
Therefore, radio observations of a sample of BALQSOs can in principle be used to 
distinguish between the evolution and orientation based models of BALs.  
Until recently such studies were practically impossible as the number of 
known radio loud BALQSOs was very small.  But the situation has dramatically 
changed with the availability of huge spectroscopic samples of QSOs  from  
Sloan Digital Sky Survey (SDSS).
For example, by stacking radio images of BALQSOs from SDSS DR3, \citet{white07} have shown that the  
BALQSOs have consistently higher radio flux densities than the non-BALQSOs.  
This may suggest that BALs are observed closer to the jet axis and is at odds with 
conventional orientation based models that require viewing angles closer to edge-on for BALQSOs. 
Studies involving radio spectral indices of BALQSOs have also been used to 
investigate the  orientation of BALQSOs. These studies revealed that BALs 
occur  
both in equatorial and polar configurations i.e. there is no preferred orientation \citep{jiang03,gregg06, ghosh}. 

Further radio observations of BALQSOs suggested that these sources are similar to compact 
steep spectrum sources (CSS; sizes$<$15\,kpc) that are considered to be the  young radio 
sources \citep{montes09,depompeo12}. 
But it is now known that in VLBI (Very Large Baseline Interferometry) i.e. milliarcsec (mas) scale resolution imaging studies, the 
radio emission associated with BALQSOs is not necessarily unresolved or compact 
\citep{jiang03,kunert07,liu08,montes09,gawronski11,bruni12,bruni13,kunert15}. 
Thus, all these studies at radio wavelengths based on image stacking, spectral indices and 
VLBI imaging techniques suggest the presence of BAL outflows at 
various orientations and in various phases of the QSO evolution.  

There have also been studies of variability of BALs in radio loud QSOs to understand 
the origin of outflows.  \citet{welling14} investigated the \civ\ BAL variability using 
a sample of 46 radio loud QSOs and reported no correlations between BAL variability and 
radio properties. However, they noted that the amplitude of BAL variability 
is typically lower for radio loud sample compared to a radio quiet sample.  
Also, they found that there is a mild tendency for the lobe-dominated QSOs to show 
greater fractional BAL variability.  These results suggest a relationship between 
the presence of radio jet and the BAL variability.
In this context, it is interesting to ask how prevalent is the jet-cloud interaction 
in BALQSOs and if it also plays a major role in driving the observed absorption line variability.

The extreme cases of absorption line variability are the ones where the flow emerges afresh or 
shows strong dynamical evolution (i.e. variation in the absorption profile and signatures of acceleration). 
Probing these emerging/disappearing BAL components (denoted as ``{\it transient BAL components}'' from now on for convenience)  can shed more light on to the hitherto unknown QSO outflow driving 
mechanisms. \citet{vivek12} reported the first emerging \mgii\ BAL in SDSS J1333+0012 and \citet{vivek12a} reported the 
disappearance of a  fine structure line BAL  in SDSS J2215-0045. \citet{filiz12} searched the SDSS-III catalog for 
such sources and reported 21 cases of \civ\ BAL trough disappearance in 19 sources. These together with previous 
reports on \civ\ BAL transients  \citep{ma02,hamann08,leighly09,Krongold10,hidalgo11} attribute them to a multiple streaming wind moving across the line of sight \citep{proga12}. 
However, most of the reported QSOs showing transient BAL absorption in the past
happen to be radio quiet. { This may be because past studies most often focused on radio quiet quasars for variability studies.}  

The objective of the present study is to understand the role played by radio jets in driving BAL outflows using 
a sample of radio { detected} BALQSOs suitable for identifying transient BAL 
absorption component. 
The sample is based on optical spectroscopic data available from the SDSS and 1.4\,GHz flux densities 
available from the Faint Images of the Radio Sky at Twenty-cm (FIRST)  survey. 
The  FIRST radio survey provides a map of the sky at 1.4\,GHz 
with a beam size of 5.4" and an rms sensitivity $\sim$ 0.15 mJy per beam. 
This manuscript is arranged as following. Section 2  describes the  radio { detected} BALQSO 
sample  used in this study. Section 3  describes the measurements of optical and radio properties of the sources and \civ\ 
broad absorption lines and Section 4 describes the analysis of the variability of BALs. Section 5  presents the statistical study of 
BALQSOs in our sample. Discussion and summary are given in Section 6. In this  work,  we  assume  a  cosmology 
with H$_0$ = 70 km s$^{-1}$ Mpc$^{-1}$, $\Omega_{M}$ = 0.3 and $\Omega_{\Lambda}$ = 0.7. 

\section{ Radio detected sample of BALQSOs}
In this study, we have used spectra of BALQSOs from the  SDSS Data 
Release-10 \citep[DR10;][]{paris14}.  In addition to the previous data 
releases, SDSS DR10 consists of data from the recently concluded Baryon 
Oscillation Spectroscopic Survey \citep[BOSS;][]{dawson13}.
BOSS used the upgraded spectrographs with larger number of fibers per plate for obtaining the spectra. The fibers used for BOSS have better throughput 
and the spectra have wider 
wavelength coverage (i.e 3600-10,400 \AA) as compared to those in 
SDSS I/II (i.e 3800-9200 \AA). 

Our initial sample consists of all the BALQSO's in the SDSS DR10 that were observed with the 
ancillary target bit name ``VARBAL'' describing photometrically-selected candidate BALQSOs.  
This sample of optically bright sources, with  i-psf  magnitude less than 19.28 mag, containing 
at least one moderately strong absorption in their BAL troughs, were primarily targeted to 
investigate BAL variability on multi-year timescales \citep{dawson13}.  We retrieved 684 
BALQSOs which have multiple epoch observations in SDSS DR10. 
Among these, 63 sources are radio detected as per the FIRST survey catalog with a typical detection limit of 1 mJy. From these 63, we discarded six  sources\footnote{J0040+0059, J0733+4625, J1349+3823, J1357+0055, J1509-0133, J1626+3752} 
where the BAL presence is 
found to be ambiguous based on our manual check. 
Seven low-$z$ QSOs\footnote{J0944+0625, J1149+3933, J1259+1213, J1324+0320, J1400-0129, J1523+3914, J1618+4113}  
contain only \mgii\ BALs and we do not include  them in the final sample.
Our final sample consists of 50 FIRST radio detected BALQSO sources containing \civ\ BAL profiles with repeat observations in the SDSS. Table~\ref{tab_lista} lists the sources in our sample with their name, Right Ascension, Declination, emission redshift, 
flux density at 1.4 GHz from FIRST survey, corresponding radio luminosity, numbers of multi-epoch spectra available in SDSS, number of \civ\ absorption components contributing to the BAL absorption trough (see the next section for our definition of \civ\ components) and an `ID'  characterising variability in \civ\ BALs. Some useful notes are also provided in the last column of the table
for individual sources. 

We also obtained the continuum light curve measurements from the Catalina Real-Time Transient Survey  \citep[CRTS;][]{drake09} for 48 sources in our final sample of 50 sources.   CRTS operates with an unfiltered set up and the resulting magnitudes are converted to V magnitudes using the transformation equation, V = V$_{ins}$ + a(V) + b(V)*(B - V), where V$_{ins}$ is the observed open magnitude, and a(V) and b(V) are the zero-point and the slope, respectively. The zero-point and slope are obtained from three or more comparison stars in the same field with the zero-point being of the order of 0.08 mag. The CRTS provides four such observations taken 10 min apart on a given night. Since we are mainly interested in the long-term variability, we have averaged these four points (or less if the measurement in one or more cases when the image of our object of interest is affected by some observational artifacts) to get the light curves.

\section{Optical continuum and absorption line properties}
For all the sources in our sample, the optical properties based on absorption lines were measured from the SDSS spectra  and the continuum variability properties were measured from the CRTS light curves.

We fitted all the spectra with a  continuum approximated by a second order polynomial for absorption/emission free regions and Gaussians for emission lines.  As we are interested mainly in studying the \civ\ BALs, we fitted the continuum only over a small rest wavelength range (1300 $<$$ \lambda_{rest}(\AA) $ $<$ 1700) relevant for the  \civ\ and  \siiv\ broad absorption lines.  
The fitting procedure involves masking the wavelength range of  absorption and bad pixels, and fitting the flux in the remaining wavelength ranges with a second order polynomial and Gaussian emission lines. We used the resulting continuum to normalize the corresponding spectrum. 
\begin{figure}
 \centering
\psfig{figure=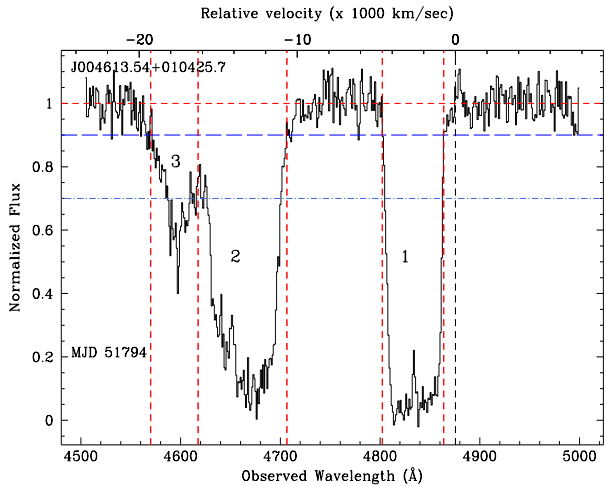,width=1.0\linewidth,height=0.8\linewidth,angle=270}
\caption{ An example demonstrating how individual \civ\ components are automatically identified in our analysis. The above plot is for the source  SDSS J0046$+$0104. Our automatic procedure first involves identifying the absorption edges where the normalized flux falls below 0.9 for 5 consecutive pixels. We further divide the identified components in to sub-components if the normalized flux rises above 0.7 continuously for 5 pixels. The velocity scale shown in the top (x-axis) is defined with respect to the \civ\ emission redshift, z$_{em}$ = 2.1551. Red dashed vertical lines mark the absorption edges of the components identified.}
\label{example_plot}
\end{figure}

We then measured the \civ\ absorption line equivalent widths from the 
normalized spectra.  This involves identifying the absorption edges and individual components. 
First, we visually inspected all the spectra to identify  BALQSOs with  at least one transient component. When a transient BAL component is noticed, we chose spectrum having the maximum absorption strength to identify the component structure. In the remaining cases, we have used the spectrum with the best signal to noise for this purpose.
 For all the spectra, wavelengths  were converted to relative velocities with respect to the \civ\ emission redshift. Absorption edges were identified as velocities where the normalized flux falls below 0.9 continuously for 5 pixels. After this first pass, the  identified components are again divided into further components if the flux rises above 0.7 continuously for 5 pixels. In this case, the mean velocity of the pixels with normalized flux values above 0.7 is taken as the absorption edge.   After identifying different components and their velocity edges, we  measure the equivalent widths for each components.  Fig.~\ref{example_plot} demonstrates this procedure for the source SDSS J0046$+$0104. The first pass identifies only two components and the second pass further divides the second component into two.   Apart from the equivalent widths, we also measure the mean depth of the BAL components to characterise the strength of the absorption. Mean depth is measured by averaging the normalized flux of three pixels around the optical depth weighted velocity centroid.       

\begin{table*}
\caption{Details of radio detected BALQSOs with multiple epoch spectroscopic observations in SDSS.   }
\begin{tabular}{|c|c|c|c|c|c|c|c|c|c|}
\hline
\hline
No. &Name		&	RA          &        Dec       &     z$_{em}$ & F$_{int}^{FIRST}$ &log(L$_{radio})$& No. of	 &No. of	& Comments$^{a}$ \\ 
    &			&	(h:m:s)	    &	(deg:m:s)      &	      &    (mJy)	   &(ergs s$^{-1}$Hz$^{-1}$)& Epochs	 &Components	&	\\
\hline                                                                                               		  		  
\hline                                                                                               		  		  
1  & SDSS J0014$-$0107 &	00:14:38.28 &     -01:07:50.19 &      1.8179 &     1.4	   &32.51 &	3	 &	2	 & NV 	 		 \\
2  & SDSS J0041$+$0017 &	00:41:18.60 &      00:17:42.49 &      1.7667 &     0.9	   &32.28 &	5	 &	1	 & V, red 	 \\
3  & SDSS J0046$+$0104 &	00:46:13.53 &      01:04:25.71 &      2.1551 &     3.0	   &33.02 &	5	 &	3	 & V, Disappearance	 \\
4  & SDSS J0053$-$0003 &	00:53:55.15 &     -00:03:09.35 &      1.7033 &     0.8	   &32.19 &	3	 &	4	 & V, Odv	 \\		
5  & SDSS J0148$-$0051 &	01:48:12.81 &     -00:51:08.78 &      1.8189 &     3.2	   &32.86 &	2	 &	4	 & V, blue    	 \\
6  & SDSS J0200$-$0845 &	02:00:22.00 &     -08:45:12.09 &      1.9423 &     7.8	   &33.32 &	2	 &	3	 & NV    	 \\
7  & SDSS J0242$+$0104 &	02:42:24.02 &      01:04:52.57 &      2.4415 &     2.6	   &33.09 &	5	 &	1	 & NV    	 \\
8  & SDSS J0743$+$3109 &	07:43:34.55 &      31:09:06.11 &      1.8983 &     1.7	   &32.64 &	2	 &	3	 & V, Odv  	 \\
9  & SDSS J0803$+$5003 &	08:03:51.59 &      50:03:17.64 &      2.9472 &     13.4	   &33.99 &	4	 &	3	 & NV	  	 \\
10 & SDSS J0811$+$5007 &	08:11:02.88 &      50:07:24.60 &      1.8394 &     23.0	   &33.73 &	5	 &	1	 & V, Disappearance	 \\
11 & SDSS J0828$+$4452 &	08:28:04.55 &      44:52:56.99 &      2.9041 &     2.4	   &33.23 &	2	 &	1	 & NV  	 \\
12 & SDSS J0835$+$4352 &	08:35:25.92 &      43:52:11.28 &      1.8218 &     1.7	   &32.59 &	2	 &	1	 & V, blue   	 \\
13 & SDSS J0855$+$0809 &	08:55:28.08 &      08:09:36.10 &      1.7696 &     8.4	   &33.25 &	2	 &	1	 & NV  	 \\
14 & SDSS J0929$+$3757 &	09:29:13.91 &      37:57:42.83 &      1.9021 &     43.1	   &34.04 &	2	 &	1	 & NV  	 \\
15 & SDSS J0945$+$5055 &	09:45:13.91 &      50:55:21.71 &      2.1331 &     2.0	   &32.83 &	2	 &	1	 & NV  	 \\
16 & SDSS J0956$+$4513 &	09:56:57.12 &      45:13:10.20 &      2.4091 &     1.7	   &32.89 &	2	 &	2	 & NV  	 \\
17 & SDSS J0959$+$6333 &	09:59:30.00 &      63:33:59.76 &      1.8478 &     14.9	   &33.55 &	3	 &	2	 & V, Disappearance  \\
18 & SDSS J1024$+$0940 &	10:24:27.35 &      09:40:29.81 &      1.8387 &     1.5	   &32.55 &	3	 &	1	 & NV  	 \\
19 & SDSS J1044$+$1040 &	10:44:52.32 &      10:40:05.87 &      1.8823 &     16.4	   &33.61 &	2	 &	6	 & V, Appearance,Odv  \\
20 & SDSS J1044$+$3656 &	10:44:59.52 &      36:56:05.27 &      2.8667 &     14.6	   &34.01 &	2	 &	5	 & NV  	 \\
21 & SDSS J1105$+$1115 &	11:05:05.03 &      11:15:41.04 &      2.4531 &     2.1	   &32.99 &	2	 &	2	 & NV  	 \\
22 & SDSS J1105$+$1512 &	11:05:31.43 &      15:12:15.84 &      2.0664 &     12.1	   &33.58 &	2	 &	3	 & V, Appearance  	 \\
23 & SDSS J1113$+$0914 &	11:13:16.31 &      09:14:39.01 &      1.6738 &     1.2	   &32.35 &	2	 &	4	 & NV   	 \\
24 & SDSS J1138$+$3704 &	11:38:03.12 &      37:04:03.00 &      1.8267 &     6.5	   &33.18 &	2	 &	1	 & NV  	 \\
25 & SDSS J1149$+$3329 &	11:49:55.67 &      33:29:07.80 &      1.9124 &     1.7	   &32.64 &	2	 &	1	 & NV  	 \\
26 & SDSS J1200$+$3508 &	12:00:51.59 &      35:08:31.56 &      1.6677 &     2.0	   &32.57 &	2	 &	3	 & NV   	 \\
27 & SDSS J1217$+$3257 &	12:17:50.15 &      32:57:11.52 &      2.0420 &     1.4	   &32.63 &	2	 &	1	 & NV  	 \\
28 & SDSS J1224$+$1010 &	12:24:10.56 &      10:10:31.07 &      1.9115 &     1.1	   &32.45 &	2	 &	1	 & NV  	 \\
29 & SDSS J1303$+$0020 &	13:03:48.95 &      00:20:10.55 &      3.6487 &     1.1	   &33.13 &	2	 &	3	 & NV  	 \\
30 & SDSS J1304$+$4210 &	13:04:25.44 &      42:10:09.83 &      1.8884 &     1.5	   &32.58 &	2	 &	2	 & NV  	 \\
31 & SDSS J1318$+$1238 &	13:18:23.75 &      12:38:12.47 &      2.6304 &     2.2	   &33.09 &	2	 &	1	 & NV, No CRTS   	 \\
32 & SDSS J1323$-$0038 &	13:23:4.559 &     -00:38:56.53 &      1.8260 &     8.9	   &33.31 &	4	 &	3	 & NV  	 \\
33 & SDSS J1324$+$0133 &	13:24:07.92 &      01:33:13.21 &      2.5036 &     6.1	   &33.48 &	2	 &	1	 & V, blue  \\
34 & SDSS J1331$+$0045 &	13:31:50.40 &      00:45:18.80 &      1.8953 &     2.9	   &32.87 &	2	 &	1	 & NV  	 \\
35 & SDSS J1334$-$0123 &	13:34:28.08 &     -01:23:49.05 &      1.7941 &     3.6	   &32.90 &	2	 &	1	 & NV  	 \\
36 & SDSS J1340$+$1232 &	13:40:14.87 &      12:32:18.23 &      1.9553 &     1.8	   &32.69 &	2	 &	1	 & V, Odv  	 \\
37 & SDSS J1444$+$0033 &	14:44:34.80 &      00:33:05.35 &      2.0361 &     12.7	   &33.58 &	2	 &	2	 & V, Odv  	 \\
38 & SDSS J1450$+$1233 &	14:50:55.92 &      12:33:31.31 &      2.7399 &     1.2	   &32.87 &	2	 &	1	 & NV  	 \\
39 & SDSS J1503$+$4401 &	15:03:32.87 &      44:01:20.63 &      2.0435 &     10.0	   &33.48 &	3	 &	2	 & NV, No CRTS  	 \\
40 & SDSS J1530$+$4409 &	15:30:49.68 &      44:09:56.87 &      1.7721 &     8.4	   &33.26 &	3	 &	2	 & V, blue  \\
41 & SDSS J1532$+$4220 &	15:32:57.60 &      42:20:47.04 &      1.9595 &     1.1	   &32.48 &	2	 &	2	 & NV   	 \\
42 & SDSS J1553$+$3245 &	15:53:55.44 &      32:45:13.32 &      2.0544 &     1.6	   &32.69 &	3	 &	3	 & V     	 \\
43 & SDSS J1601$+$2947 &	16:01:38.40 &      29:47:34.44 &      1.9677 &     1.1	   &32.48 &	2	 &	2	 & NV   	 \\
44 & SDSS J1603$+$3002 &	16:03:54.23 &      30:02:08.51 &      2.8321 &     53.7	   &34.56 &	2	 &	1	 & NV   	 \\
45 & SDSS J1618$+$3301 &	16:18:12.96 &      33:01:55.91 &      2.0031 &     5.3	   &33.19 &	2	 &	1	 & NV   	 \\
46 & SDSS J1621$+$3555 &	16:21:43.68 &      35:55:33.96 &      2.0646 &     1.2	   &32.57 &	2	 &	3	 & NV   	 \\
47 & SDSS J1632$+$2201 &	16:32:39.36 &      22:01:41.87 &      1.9537 &     1.2	   &32.52 &	3	 &	3	 & NV   	 \\
48 & SDSS J1641$+$3058 &	16:41:52.32 &      30:58:51.60 &      2.7687 &     2.1	   &33.13 &	2	 &	3	 & NV   	 \\
49 & SDSS J1655$+$3945 &	16:55:43.19 &      39:45:19.80 &      1.7530 &     10.1	   &33.32 &	2	 &	2	 & V, Disappearance	 \\
50 & SDSS J2353$-$0050 &	23:53:13.68 &     -00:50:23.21 &      1.9588 &     1.2	   &32.52 &	2	 &	1	 & NV  	 \\
\hline
\hline
\end{tabular}
 \begin{flushleft}
$^{a}$ V - variable BAL; NV - Non-variable BAL; red- red part of BAL varying; blue - blue part of BAL varying; Odv - variation in optical depth; Appearance - New BAL component appeared; Disappearance - BAL component disappeared.
  \end{flushleft}
\label{tab_lista}
\end{table*}

\begin{figure*}
 \centering
\psfig{figure=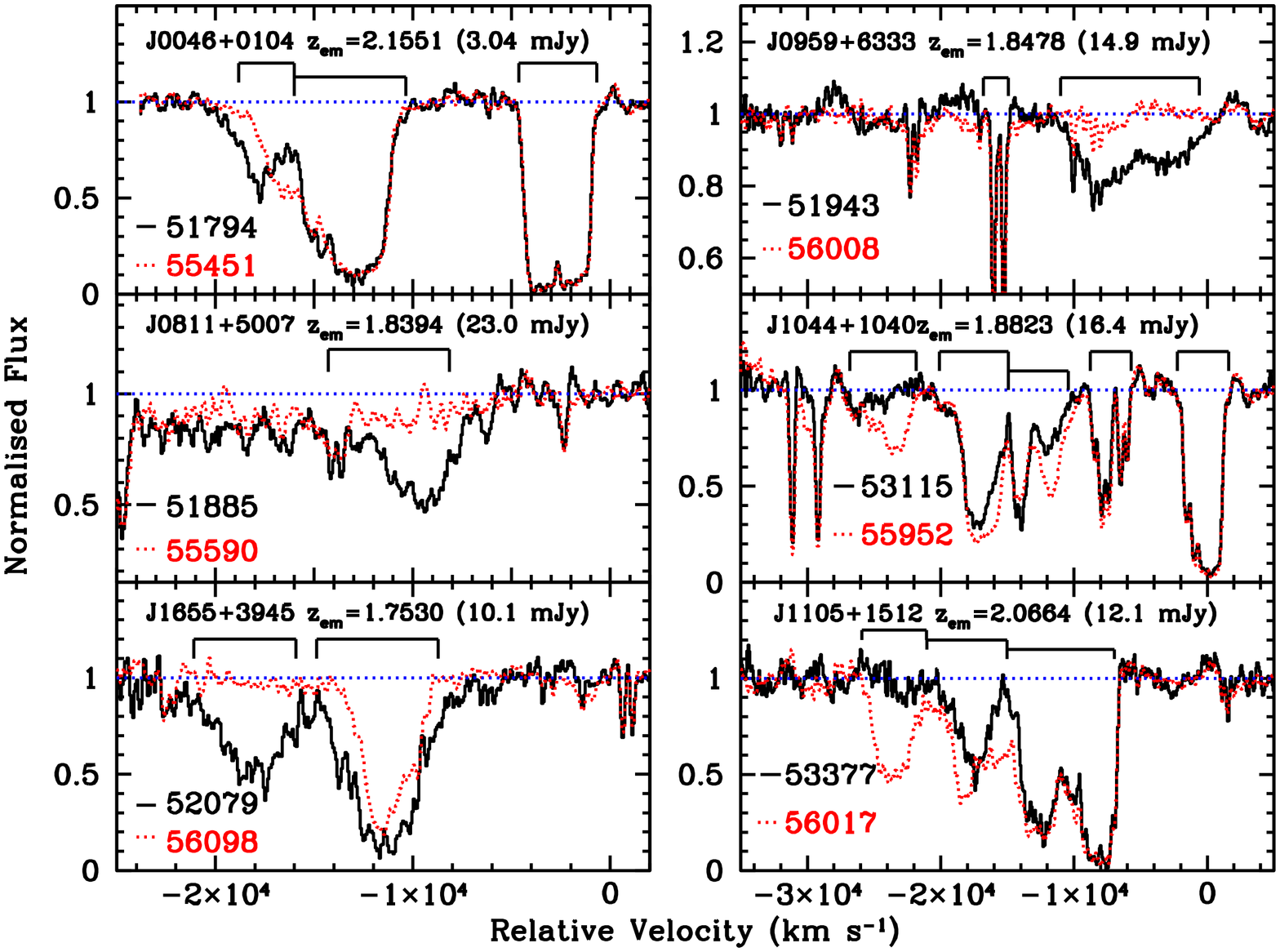,width=1.0\linewidth,height=0.8\linewidth,angle=270}
\caption{  Normalized spectra of six sources showing transient \civ\ BALs. 
The spectra shown with dotted red line corresponds to later epoch measurements. The horizontal lines mark the velocity ranges for different components, identified by the method described in Fig~\ref{example_plot}, that are used for measuring the equivalent widths. Relative velocity is measured with respect to the centroid of the \civ\ emission line. Emission redshift, epoch of observations and the 1.4 GHz radio flux density from FIRST for each source are also provided. 
 }
\label{emerge_plot}
\end{figure*} 


We also constructed a control sample of radio quiet BALQSOs from the SDSS DR10 catalogue to quantify the role played by the radio jets in  driving the transient BAL components in radio { detected} BALQSOs. For each of the six quasars exhibiting transient BAL components in our sample (see next Section), we obtained 10 radio quiet BALQSOs  matching closely in the absolute i-band magnitude and redshift.   

  \section{Variability of BALs: Analysis and Results \protect\footnote{ I\lowercase{n this work, we also refer to the} BAL QSO\lowercase{s showing transient \civ\ components as transient} BAL\lowercase{s.}}}
We visually checked all the available spectra of individual sources for \civ\ absorption line variations by overplotting  different epoch spectra.
Optical depth variability in \civ\ BALs are detected in several sources in our sample without showing large variations in the velocity profile. However, in six sources, we see dramatic variations in the BAL profiles. In the latter epochs of these sources, either a new velocity component has  emerged afresh or an already existing component has completely disappeared.

Fig.~\ref{emerge_plot} shows the  normalized spectra of these six sources showing \civ\ BAL transient components. 
Red dashed spectra correspond to later epoch measurements. Horizontal blue dashed line corresponds to the unabsorbed normalized continuum. Black horizontal lines mark the absorption edges for different components used for our equivalent width measurements. Relative ejection velocity of a component is the maximum velocity of a component  measured with
respect to the redshift of the \civ\ emission line. Redshift, epoch of observations and the peak radio flux density at 1.4 GHz for each source are also presented in each panel. 
 
SDSS J0046+0104, SDSS J0811+5007,  SDSS J0959+6333 and SDSS J1655+3945 have atleast one \civ\ BAL component which disappeared in the later SDSS epoch and  in SDSS J1044+1040 and SDSS J1105+1512 a new component is observed to have emerged in the later SDSS epoch. \citet{filiz12} has already identified the BAL disappearance in SDSS J0811+5007. Interestingly, 5 out these six transient BAL sources have peak radio flux density at 1.4 GHz greater than 10 mJy. In the following section, we perform different statistical 
tests to understand the influence of the radio emitting plasma on the \civ\ outflows. 

\section{Statistical Analysis}
	In this section, we investigate the correlations between the BAL variability and other properties of the sample.  We use the absolute  fractional variation in the \civ\ equivalent width of individual components to quantitatively characterise the BAL variability. Absolute fractional variation in the equivalent width is defined by $|\Delta W|/<W>$ where $\Delta W$ and $<W>$ are the equivalent width differences and average equivalent width between two epochs respectively. This approach is particularly useful in studying the transient 
BAL cases where $|\Delta W|/<W>$ close to 2  corresponds to a completely emerged or disappeared
BAL component. 

\begin{figure}
 \centering
\psfig{figure=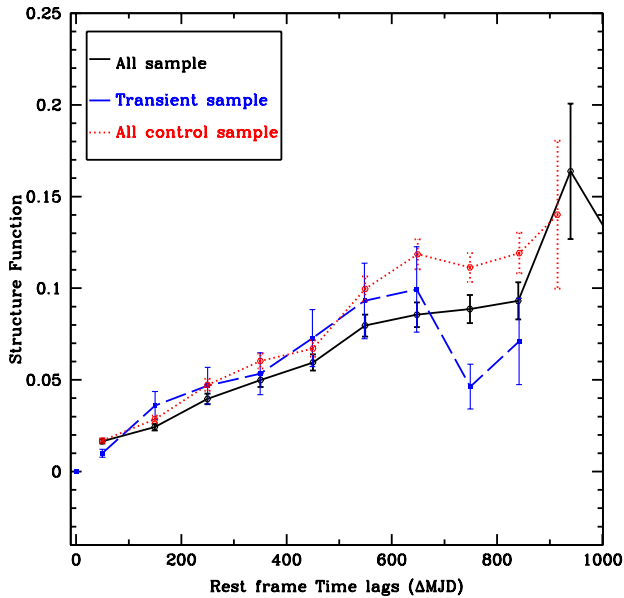,width=1.0\linewidth,height=1.0\linewidth,angle=0}
\caption{ The black (solid), blue (dashed) and red (dotted) curves represent  the structure function (in mag) of the CRTS magnitude variations of all the sources given in Table~\ref{tab_lista},  \civ\ BALs with transient components in our sample and the radio quiet control sample respectively. 
}
\label{sf_plot}
\end{figure} 

However for practical purpose, we define a BAL component to be a transient if the  absolute fractional variation in the \civ\ equivalent width is greater than unity. This is because our definition of individual component may have included some additional absorption from gas unrelated  to the transient component.  All the 6 sources showing distinct \civ\ BAL transient signatures during our visual inspection are clearly  selected  when we impose this threshold. We compare properties of this transient \civ\ BAL sub-sample with the rest of the sample to understand the BAL transient phenomena. In particular, we wish to address whether there is any common observational correlations between cases with simple optical depth variability and the occurrence of new transient components. 

\subsection{Correlation with continuum parameters}
Previous BAL variability studies have suggested that optical depth variations are not strongly correlated with continuum flux variations \citep{lundgren07,gibson10,vivek14,welling14}. In the case of \mgii\ emerging BALQSO SDSS J1333+0012, there is a suggestion that the emergence may be accompanied with continuum variations that can be interpreted as disturbances in the accretion disk causing the new emerging component \citep{vivek12}. In order to explore the role played by the continuum variations in the case of 
\civ\  BALs with transient components, it is important to compare the nature of continuum variability of these BALQSOs with the rest of the BALQSO population. 

In this section, we address this 
using the structure function analysis of CRTS light curves. The structure function (SF) has been used extensively in the literature as a variability diagnostic  for either  individual or an ensemble of QSOs \citep{vandenberk04,welsh11,macleod12}. Several varying definitions of structure function have been used in literature which essentially quantify the amplitude of variability as the function of observed time-lags \citep[See,][for various definitions of SF]{graham14}.  

For our present study, we have used the formulation of structure function  and the
associated error ($SF_{err}$) given by \citet{macleod12}:
\begin{eqnarray}
\centering
SF = 0.74*(IQR);\hspace{0.3cm}SF_{err} = SF*1.15/\sqrt{(N-1)}
\label{eq1} 
\end{eqnarray}
where IQR is the 25\%–75\% interquartile range of the $\Delta$m distribution  and N is the number of $\Delta$m values.  We chose to bin the magnitude differences within intervals of 20 rest-frame time lag days before calculating the structure function. \citet{welling14} compared the structure functions of a sample of radio loud and radio quiet BALQSOs and reported similarity in the optical continuum variability between the two samples. \citet{joshi13} also report a similarity in the microvariability (i.e., variability with in a day) properties of radio loud and radio quiet BALQSOs.

Fig.~\ref{sf_plot} shows the plot of absolute V-band magnitude variations at different time lags. The black solid, 
 blue dashed and red dotted curve represents the structure function of the CRTS magnitude variability computed using Eq.~\ref{eq1} for all the 
QSOs in the sample, the transient BALQSOs in the sample and the radio quiet control sample 
respectively. It is clear that the variability properties of the QSOs with transient BALs  and
the rest are identical. { The structure function of  BAL QSOs with transient BAL component  seems to be lower at time scales greater than 800 days.  This would mean that  BAL QSOs showing transient BAL component are less variable at longer time scales and is contrary to what is expected in the case of ionization driven outflows.} This implies that the transient BAL  phenomenon is most unlikely  due to the ionization changes. 

It is interesting to note that the SFs derived here are very much
consistent with those of \citet{welling14} (see their Fig.~14). 
They found that the optical continuum variability of radio loud and radio quiet
BALQSOs are similar in their sample. 
Therefore, we can conclude
that overall optical variability of our radio detected BALQSO is also similar to 
that of radio quiet BALQSOs.

%

\begin{figure*}
 \centering
\begin{tabular}{c c}
\psfig{figure=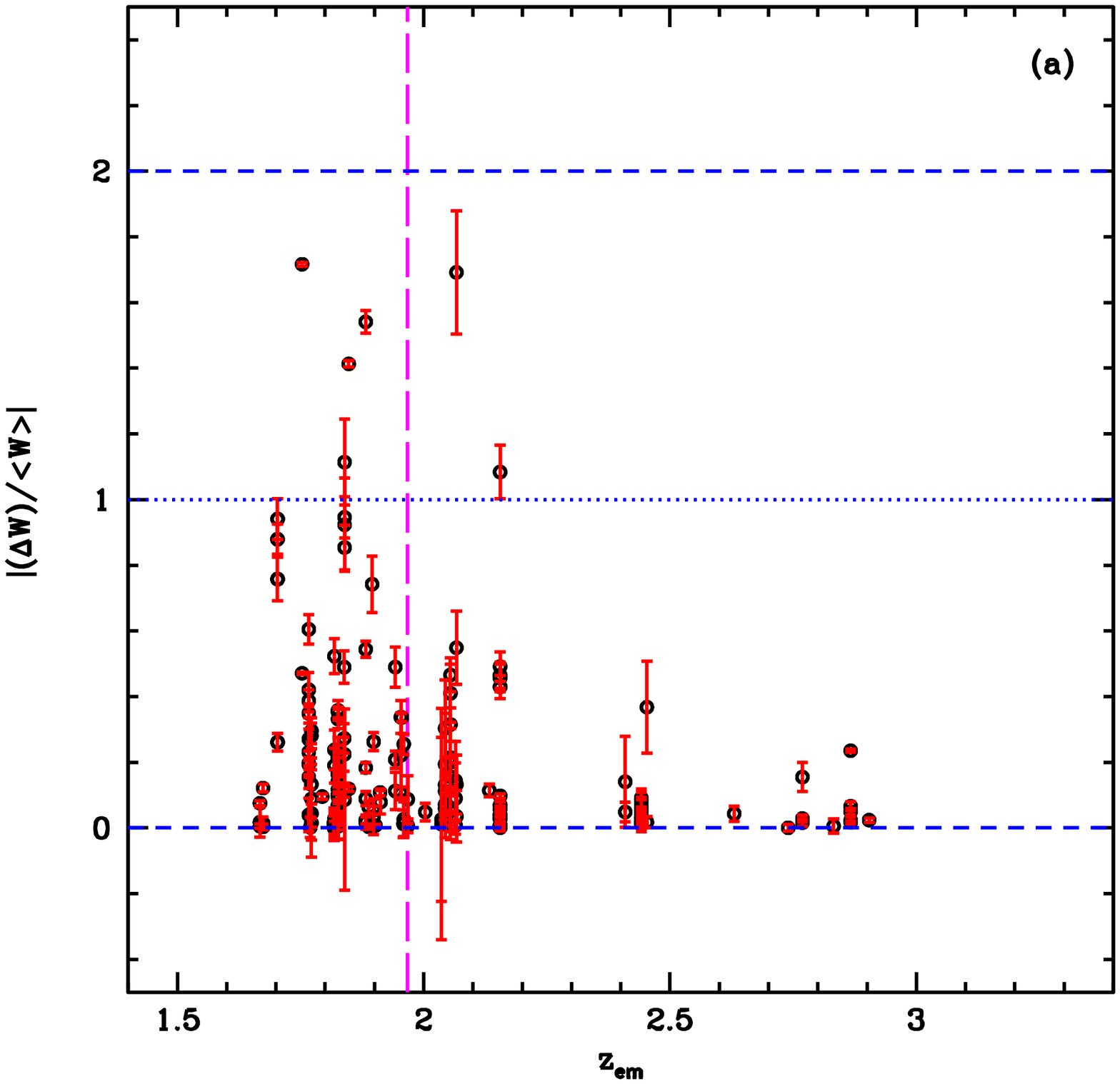,width=0.5\linewidth,height=0.35\linewidth,angle=0}&
\psfig{figure=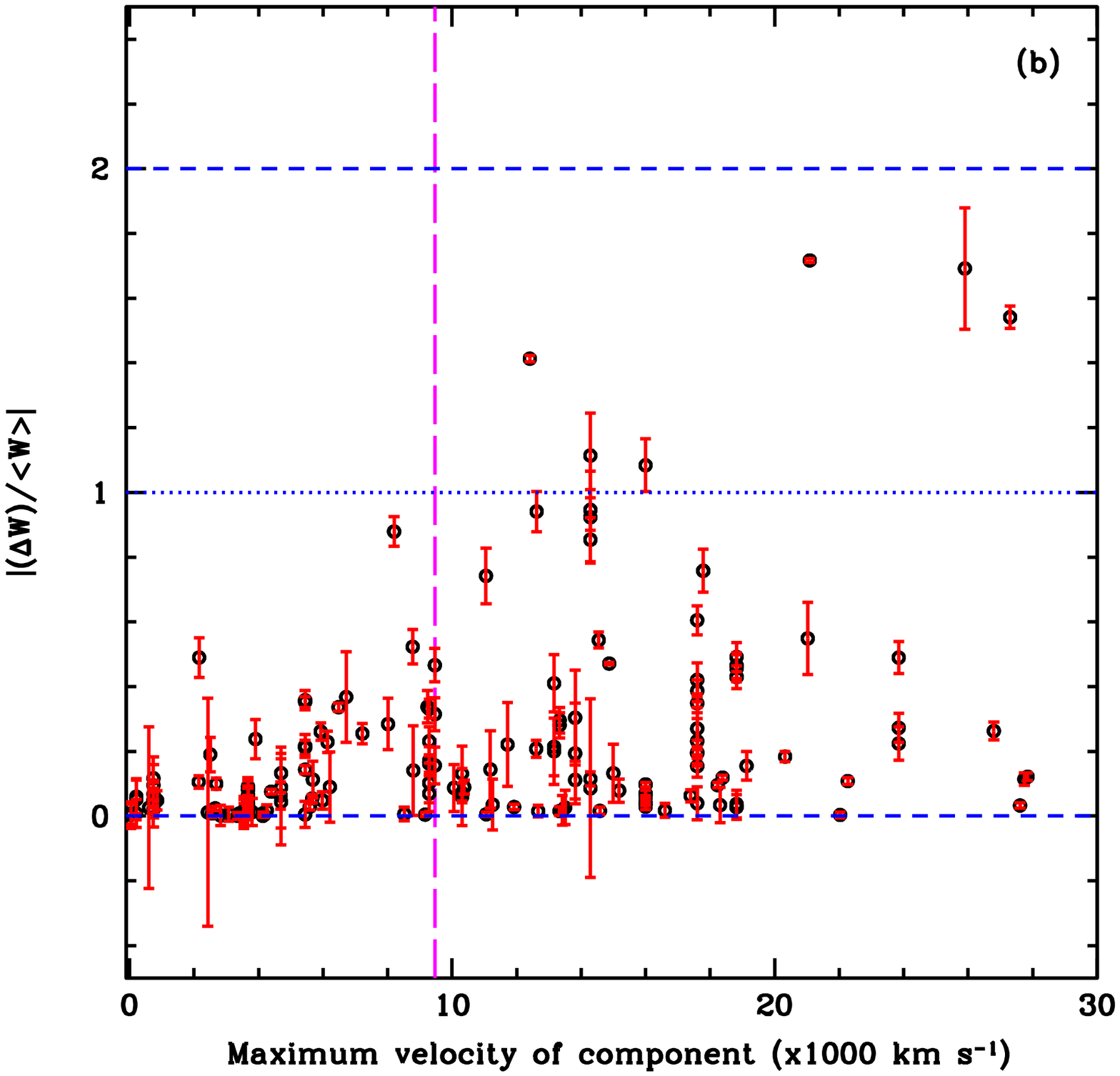,width=0.5\linewidth,height=0.35\linewidth,angle=0}\\
\psfig{figure=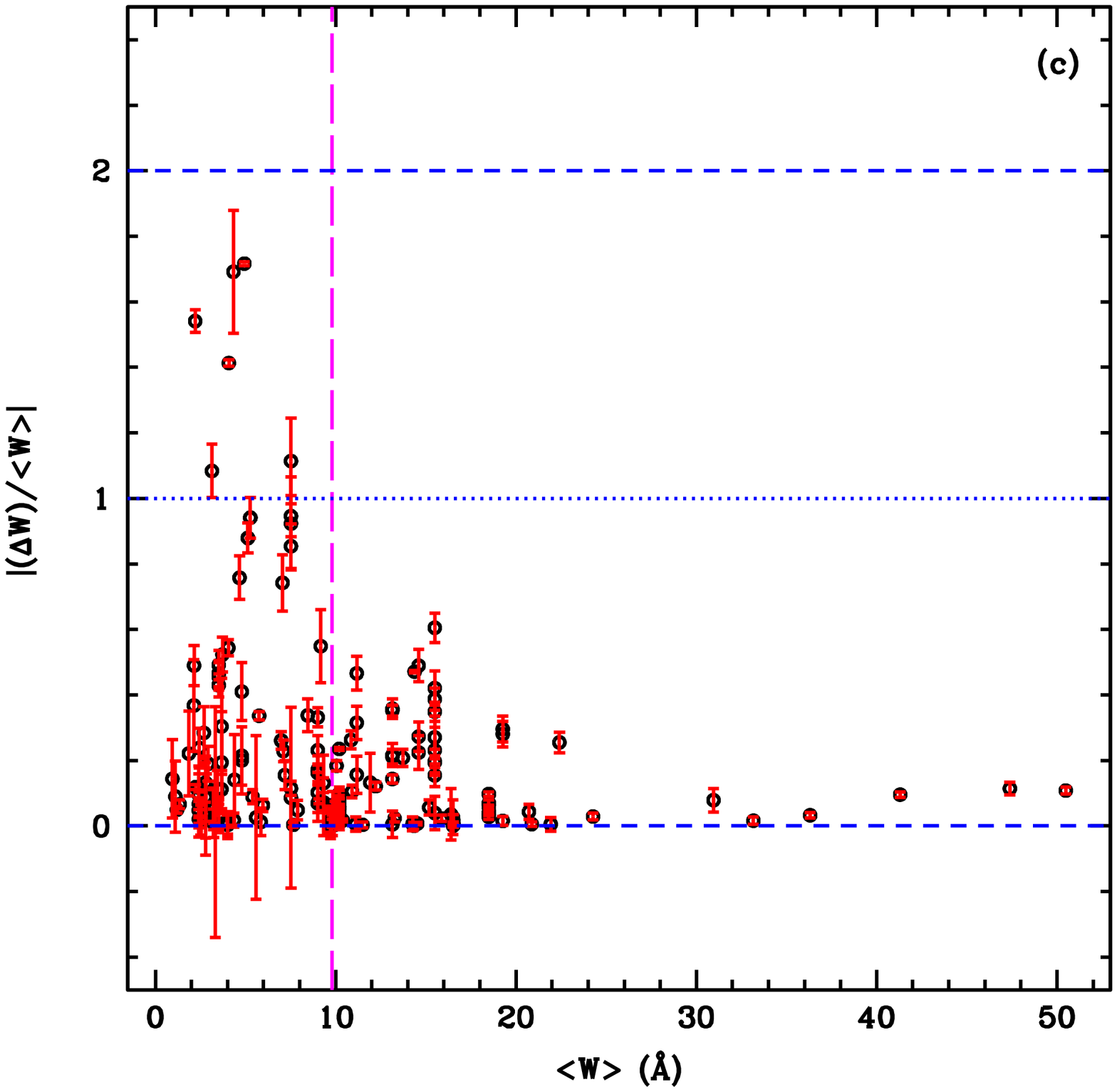,width=0.5\linewidth,height=0.35\linewidth,angle=0}&
\psfig{figure=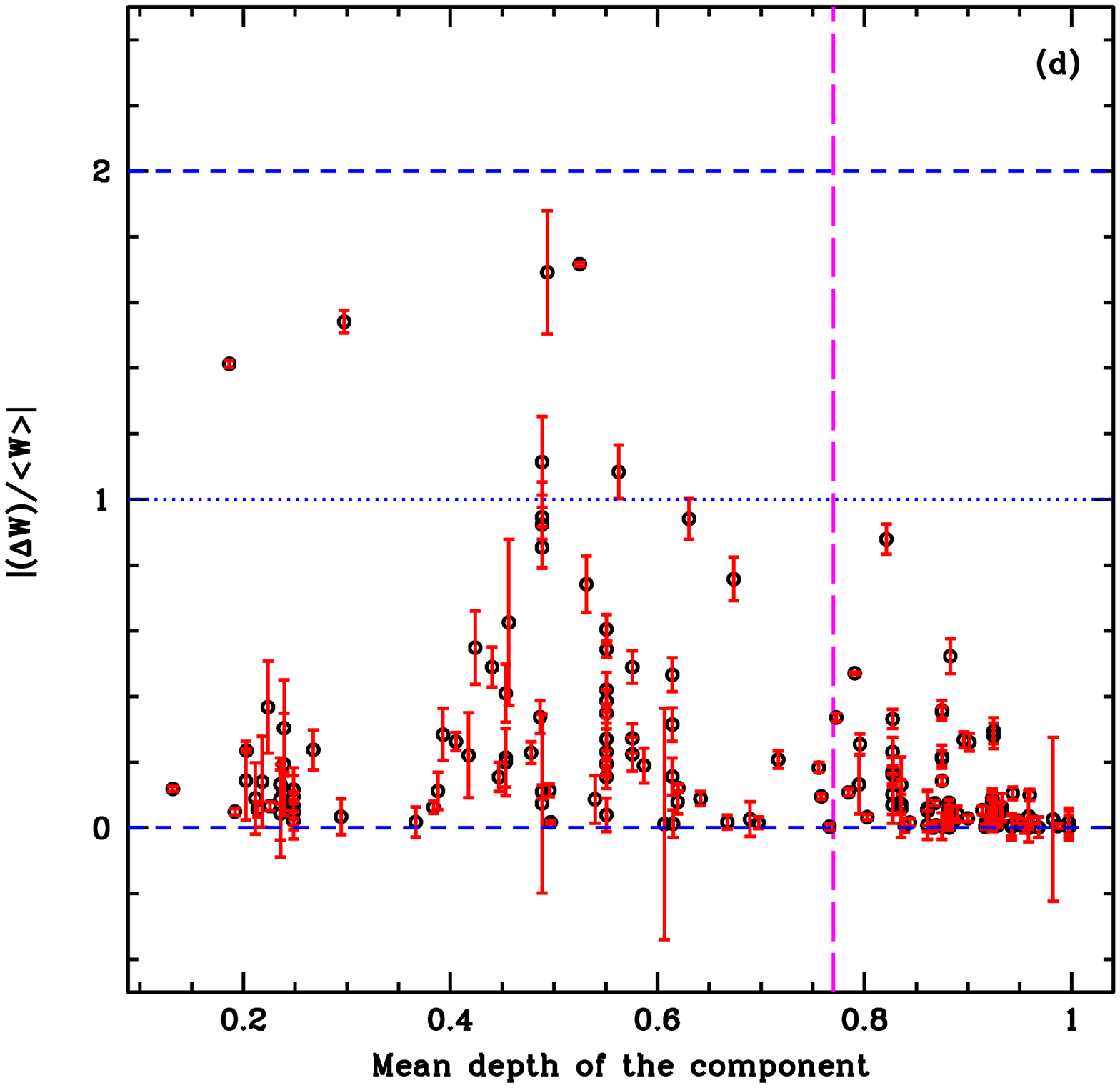,width=0.5\linewidth,height=0.35\linewidth,angle=0}\\
\psfig{figure=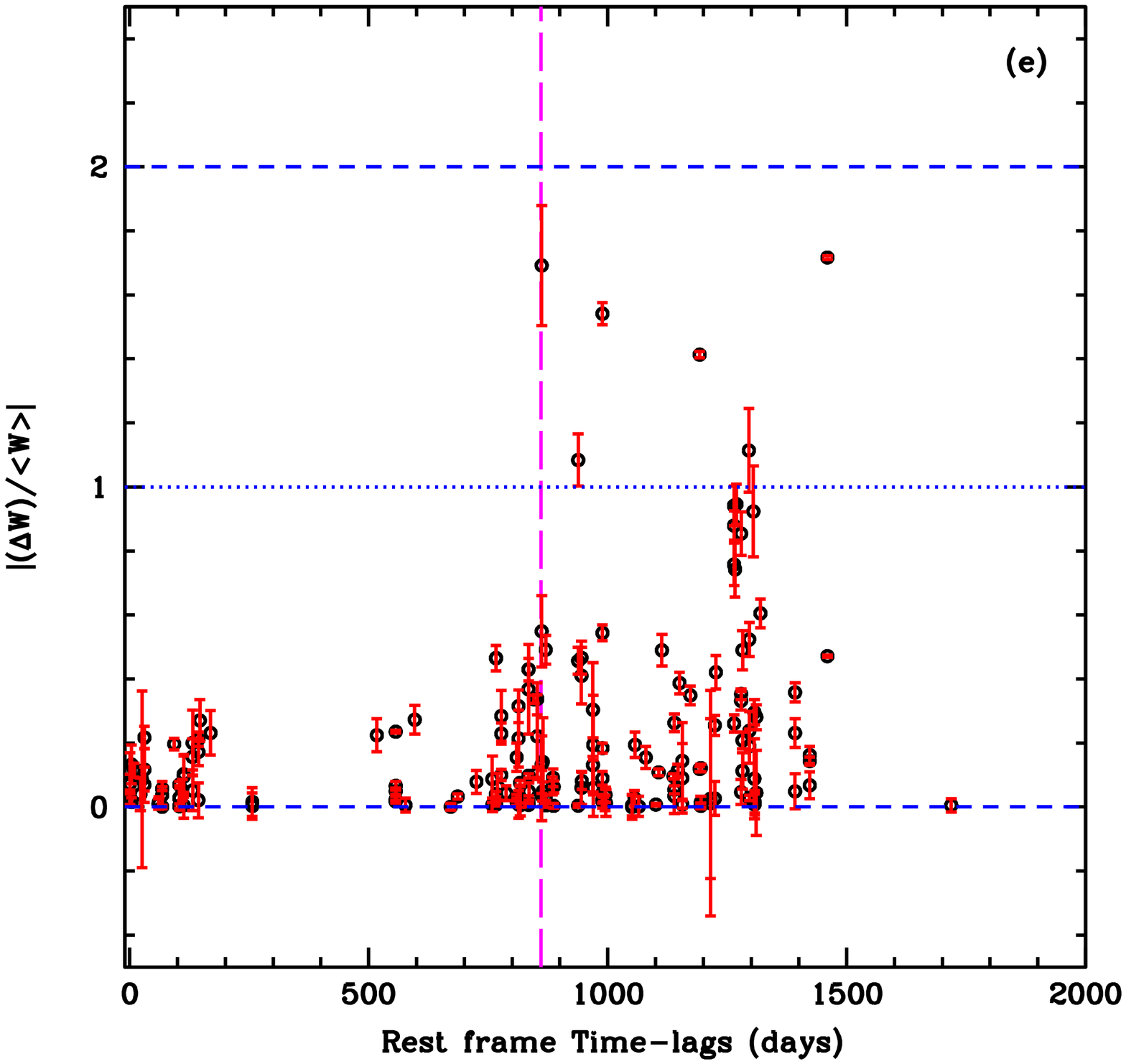,width=0.5\linewidth,height=0.35\linewidth,angle=0}&
\psfig{figure=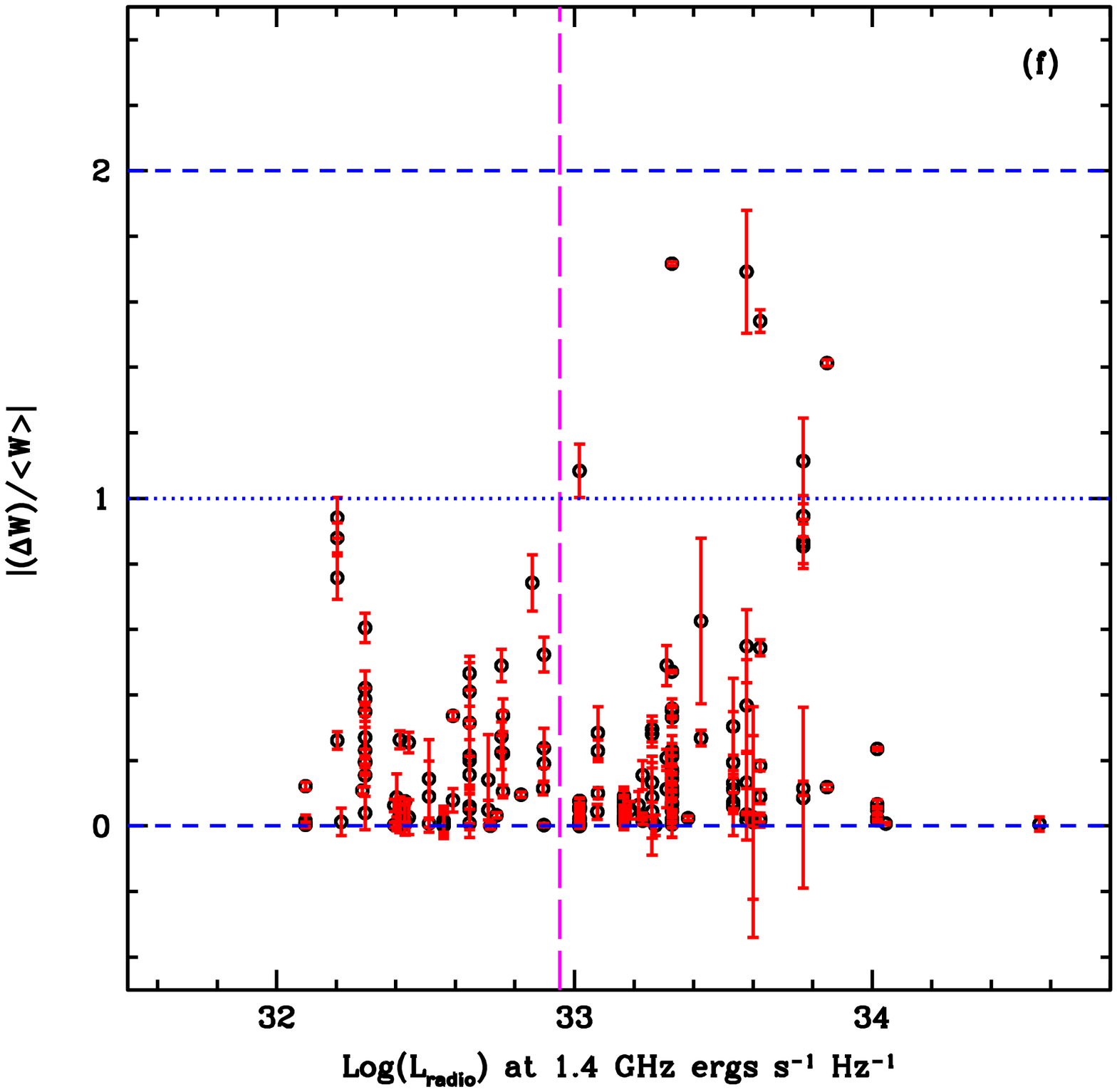,width=0.5\linewidth,height=0.35\linewidth,angle=0}\\

\end{tabular}
\caption{Absolute fractional variation of \civ\ equivalent widths of individual components are plotted against emission redshift, maximum velocity of the component, average equivalent width, mean depth, timelags and integrated radio luminosity. The upper blue horizontal dashed line represents a fractional variation of 2 which corresponds to a transient BAL component. The blue dotted horizontal line represents a fractional variation of 1. The median value of the distributions are marked by the vertical long dashed magenta line in each plot.
 }
\label{fr_W_civ}
\end{figure*}


\begin{table*}
\caption{ Results of two-sided KS test for the parent radio { detected} BALQSO sample (60 objects) and the subsample of 6 transient BALQSOs }
\begin{tabular}{|c|c|c|c|}
\hline
\hline
Parameter& D & D$_{critical}^{\dagger}$ & p-value   \\ 
  &  	    &   &	\\
\hline
\hline
Ejection velocity	&    0.40   &   0.34	&0.01	\\  
Radio Luminosity	&    0.59   &   0.59  &       0.03   \\
FIRST radio  flux	&    0.63   &   0.59	&	0.01	\\  
Redshift   		&    0.33   &   0.59	&	0.53		\\  
Timelags		&    0.18   &   0.22  &	0.17	\\ 
Bolometric Luminosity	&    0.35   &   0.59	&	0.45	\\  
Eddington Ratio		&    0.18   &   0.59	&	0.99	\\  
M$_{BH}$		&    0.15   &   0.59	&	0.99	\\  
(SDSS r - WISE W4) color&    0.53   &   0.59	&	0.07	\\  
\hline   
\hline 
 \end{tabular}
 \begin{flushleft}
$^{\dagger}$ D$_{critical}$ values evaluated for a statistical significance, $\alpha$ = 0.05 \\ 
  \end{flushleft}
\label{tab_kstest}
\end{table*}

\begin{table*}
\caption{Properties of transient BALs in the radio detected and radio quiet sample}
\begin{tabular}{|c|c|c|c|c|c|c|c|}
\hline
\hline
No. &Name	& Right Ascension & Declination  &  z$_{em}$ & i & V$_{eje}^{a}$ & $\Delta$MJD$^{b}$ \\ 
    &		&	    &&  &	(mag) &	  (\kms)  &   (days)   	\\
\hline
\multicolumn{8}{|c|}{Radio detected sample} \\                                                                                            		  
\hline  
1  & J0046+0104& 00 46 13.54&  	+01 04 25.7  &    2.1551 &   -28.39 & -18819.9 &	870.4	\\ 
2  & J0811+5007& 08 11 02.90&	+50 07 24.5  &    1.8394 &   -26.71 & -14286.8 & 	1279.1	\\
3  & J0959+6334& 09 59 29.85&	+63 34 00.2  &    1.8478 &   -29.11 & -12410.4 & 	1459.8	\\
4  & J1044+1040& 10 44 52.41&	+10 40 05.9  &    1.8823 &   -28.17 & -27299.4 & 	1192.4	\\
5  & J1105+1512& 11 05 31.41&	+15 12 15.9  &    2.0664 &   -27.35 & -25897.6 & 	975.2	\\
6  & J1655+3945& 16 55 43.23&	+39 45 19.9  &    1.7530 &   -27.45 & -21089.9 & 	861.9	\\
\hline                                            
\multicolumn{8}{|c|}{Radio quiet control sample} \\                                                                                      		  
\hline 
7  & J0119+0043& 01 19 48.52&	+00 43 55.9  &    1.7668 &   -27.36 & -2098.9  & 	1341.7	\\
8  & J0825+2607& 08 25 40.90&	+26 07 37.4  &    1.7503 &   -27.13 & -16119.9 & 	1070.3	\\
9 & J0946+3800& 09 46 02.23&	+38 00 59.3  &    2.0678 &   -27.45 & -25149.9 & 	843.2	\\
10 & J1007+0304& 10 07 16.69&	+03 04 38.7  &    2.1241 &   -27.87 & -18794.6 & 	848.4	\\
11 & J1415+3956& 14 15 33.99&	+39 56 27.3  &    1.8270 &   -26.59 & -18698.5 & 	1042.2	\\
12 & J1448+1228& 14 48 26.10&	+12 28 14.7  &    2.0694 &   -27.19 & -18987.0 & 	814.2	\\
13 & J1636+2051& 16 36 28.41&	+20 51 21.6  &    1.7411 &   -27.19 & -13735.3 & 	813.1	\\

\hline   
\hline 
 \end{tabular}
 \begin{flushleft}
$^{a}$ V$_{eje}$ is the ejection velocity of the transient component.\\
$^{b}$ shortest rest-frame timescale over which the BAL transience observed.\\
  \end{flushleft}
\label{tab_emlist}
\end{table*}

\subsection{BAL transience and absorption line parameters}

Previous  BAL variability studies have found positive correlations of BAL variability 
with observed time-lag and ejection velocity, and a negative correlation with average equivalent 
width \citep{lundgren07,gibson08,gibson10,capellupo11,filiz13,vivek14}. In this section,
we study the relationship between these quantities and transient \civ\ BAL phenomenon. 

Fig.~\ref{fr_W_civ} shows the absolute fractional variation of \civ\ equivalent widths 
in individual absorption components with z$_{em}$, maximum velocity of the component, average \civ\ equivalent widths,  
mean depths of BAL components, rest-frame time lags and integrated radio  luminosity.  In each panel, the blue upper dashed and  dotted  horizontal  lines 
represent a fractional variation  of 2 and 1 respectively. The median value of  the quantity in the 
abscissa is marked by the dashed magenta vertical line in each panel. Clearly there are six 
cases where the equivalent width fractional variation are between 2 and 1. In two cases, SDSS J0046+0104 and SDSS J0811+5007, the equivalent width fractional variation is close to 1. In the case of SDSS J0046+0104, the transient BAL component is weak and in the case of SDSS J0811+5007, the fractional variation values are suppressed by the inclusion of a small component in the high velocity end by our automatic algorithm. Panel (a) of Fig.~\ref{fr_W_civ} shows that there is no dependence of the occurrence of 
transient BAL on redshift.  
This is also confirmed by the two-sided Kolomogrov-Smirnov (KS) test results summarized in Table.~\ref{tab_kstest}.  

From panel (b) in Fig.~\ref{fr_W_civ}, it is clear that the fractional equivalent width change shows a large scatter at high maximum velocities.
In particular absorption components that show change in fractional equivalent width in excess of 0.8 have ejection velocity in excess of
8000 \kms. All the 6 components identified as transient BALs have ejection velocities in excess of 11,000 \kms. This confirms that 
transient BAL phenomenon is more common among components with large ejection velocities. 
This is also confirmed by the KS-test results
summarized in Table~\ref{tab_kstest}.

In panel (c) in Fig.~\ref{fr_W_civ} we plot fractional \civ\ equivalent width as a function of mean \civ\ equivalent width. It is clear from the figure that the transient
components tend to have the average \civ\ equivalent width less than 8 \AA\ and mean depth of the \civ\ absorption trough less than 0.55 (see panel (d)
of Fig.~\ref{fr_W_civ}). This is consistent with the transient phenomenon occurring mainly in components that are typically shallow and have low equivalent widths.
The low \civ\ equivalent width could either mean low covering factor or unsaturated absorption line.

It is clear from panel (e) in Fig.~\ref{fr_W_civ} that the envelope of fractional \civ\ equivalent width variations widens at long observed time-lags.  
In particular all the transient BAL components are seen at rest-frame timescales of more than 800 days. 

In summary, we find that the transient \civ\ BAL phenomenon is more frequent in components with large ejection velocities having smaller \civ\ equivalent widths and
monitored over a timescale of more than 800 days in the rest frame of the QSOs. This is similar to what was found
by \citet{filiz12} for disappearing \civ\ absorbers predominantly towards radio quiet QSOs.
 Interestingly the above found trend for the
transient \civ\ BAL components are also seen for the general optical depth variability. 
Therefore, purely based on the discussions
presented here the transient \civ\ BAL components may appear to be the extreme case of \civ\ variability
\citep[see][]{filiz13}. In the following section, we investigate whether the frequency of occurrence of 
the transient \civ\ BAL components is significantly higher in the radio { detected} QSOs.

\begin{figure}
 \centering
\psfig{figure=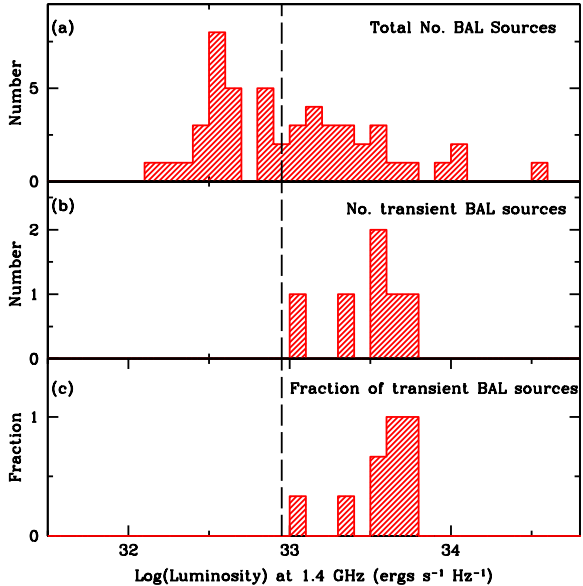,width=1.0\linewidth,height=1.0\linewidth,angle=0}
\caption{ Radio luminosity distribution of \civ\ BALQSOs.  The panels (a), (b) and (c) show the distribution for all the BALQSOs, transient BALQSOs and the fraction of transient BALQSOs, respectively. Black vertical dashed line corresponds to the median radio flux density of the sample.  
}

\label{hist_fr_pf}
\end{figure}

\begin{figure}
 \centering
\psfig{figure=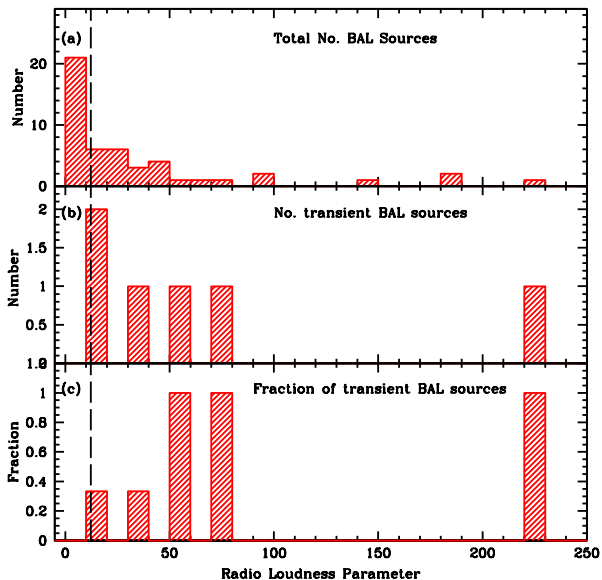,width=1.0\linewidth,height=1.0\linewidth,angle=0}
\caption{ 
Radio loudness parameter distribution of \civ\ BALQSOs.  The panels (a), (b) and (c) show the distribution for all the BALQSOs, transient BALQSOs and the fraction of transient BALQSOs, respectively. Black vertical dashed line corresponds to the median radio loudness parameter of the sample.   
}
\label{hist_rad_loud}
\end{figure}

\begin{figure}
 \centering
\psfig{figure=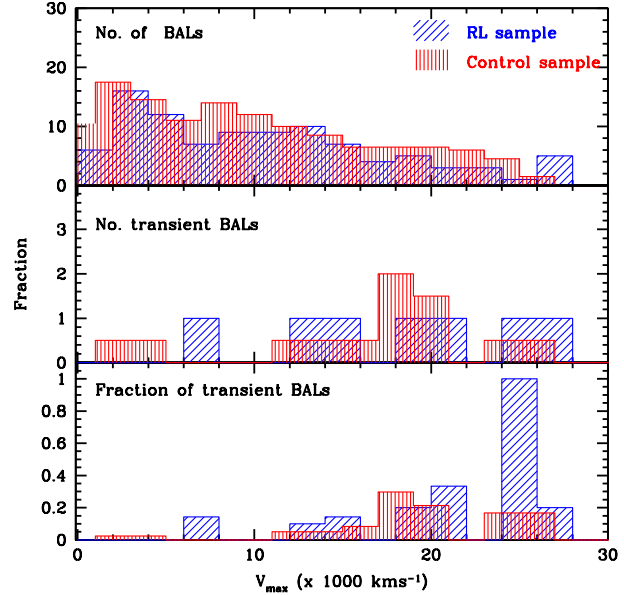,width=1.0\linewidth,height=1.0\linewidth,angle=0}
\caption{Histogram distribution of maximum ejection velocities of all the \civ\ BAL components. The blue (with slanted lines) and red (with vertical lines) histograms corresponds to the radio detected and radio quiet control sample respectively. {\it Top panel} : number of \civ\ BAL absorption components at each ejection velocity. {\it Middle panel } : number of transient \civ\ BAL components. {\it Lower panel} : the fraction of transient \civ\ BAL components .    }

\label{hist_vel}
\end{figure}

\subsection{BAL transience and radio emission}

We first noticed that the transient \civ\ BALs occur in sources with high integrated radio fluxes. 
While the two BALQSOs \footnote{ SDSS J0929+3757 and SDSS J1603+3002 have  radio fluxes 43 mJy and 53 mJy respectively. In both these cases, \civ\ BALs are defined by one strong component (probably saturated) at low ejection velocities. } with high radio flux density does not show a transient \civ\ absorption component, we find a tentative evidence for high fraction of transient \civ\ BAL components  
in BALQSO with flux density in excess of 10 mJy. 
 To better characterize the intrinsic radio emission, we converted the integrated radio fluxes to luminosities. In panel (f) of Fig.~\ref{fr_W_civ} we plot the fractional equivalent width variations as a function of radio luminosity. It is apparent
from this figure that the transient \civ\ BAL components predominantly occur in radio bright sources.
The KS-test results presented in Table~\ref{tab_kstest} confirms that the radio flux/luminosity distributions in BALQSOs with transient \civ\ absorption
are different from the rest of the QSOs.

This trend is also apparent in Fig.~\ref{hist_fr_pf}.  
In this figure, 
the top panel (a) shows the radio luminosity distribution of  BALQSOs. 
The red (shaded)  histogram in panel (b) shows the distribution of QSOs with a transient \civ\ BAL. 
Similarly, panel (c) shows the fraction of QSOs with transient \civ\ BALs in red.  
If we divide the sample into two at the median radio luminosity value 
(i.e, Log(L$_{radio}$) = 32.9 ergs s$^{-1}$ Hz$^{-1}$), 25$\pm$10 \% of sources with  radio  luminosity greater than the median  show transient \civ\ BALs whereas 
none of the sources  with  radio luminosity less than the median show transient \civ\ BAL. 
Thus we have a tentative evidence (at 2.5 $\sigma$ level) for the transient \civ\ BALs to be more probable 
in radio bright QSOs. Two sample  KS tests between the  transient and the parent sample shows that these two samples are different by 99\%, 99\% and 97\% in their ejection 
velocity, radio flux density and radio luminosity distributions (see Table~\ref{tab_kstest}).  
Fig.~\ref{hist_rad_loud} shows the distribution of radio loudness for BALQSOs. 
Clearly, all the  \civ\ BALs with a transient component have radio loudness parameter greater than the median of 
the radio loudness parameter distribution, further confirming the result.

 To explore this further, we consider the radio loud BAL sample of \citet{welling14}. 
They estimated the 5\,GHz radio luminosity for their sample using FIRST peak fluxes for the core, and 
integrated fluxes for the lobes, assuming radio spectral indices of $\alpha$=$-$0.3 and $-$0.9, respectively. 
To make a straight-forward comparison with our sample, we consider the compact (unresolved in FIRST) BALQSOs in 
their sample having the  5\,GHz radio luminosity (in log) $>$ 32.7.  For $\alpha$=$-$0.3 this corresponds to a 
log(L$_{radio}$) = 32.9 at 1.4\,GHz i.e. the median value for our sample. Thus, considering only compact 
BALQSOs in the two samples with  log(L$_{radio}$) $\ge$ 32.9 at 1.4\,GHz, we find that the fraction of 
transient BALs in our and their samples are 27\% and 10\% respectively. While the difference could simply be due to the 
small number statistics, we notice that not all QSOs in the sample of \citet{welling14} are monitored for $>$ 800 days i.e. 
 the timescale over which most of the transient \civ\ absorption occur in our sample. For example, only 5 out of 21 compact BALQSOs with log(L$_{radio}$) = 32.9 at 1.4\,GHz in the 
their sample are monitored over a rest frame time-scale of greater than 800 days. The two sources showing possible transience
are among them i.e. a detection rate of 40$\pm$28 \%.  { In our radio detected sample, there are 20 sources with log(L$_{radio}$) $\ge$ 32.9 at 1.4\,GHz and rest-frame timescales $\ge$ 800 days. Six  out of this 20 sources show transient nature (i.e., detection rate of 30$\pm$12 \%). Thus, with in statistical uncertainties, there is no major discrepancy in the transient BAL detection rate between our sample and the sample of \citet{welling14} as long as we confine to similar monitoring periods. This exercise also establishes the importance of long time-scale monitoring in addition to having large enough sample to have statistically significant results.}


To further investigate the finding of `higher' fraction of transient \civ\ BALs at high radio flux densities/luminosities in our sample, we  
proceed to study the fraction of transient \civ\ BAL sources in radio quiet BALQSOs. A clear deviation from the radio detected sample will firmly establish the role played by the radio jets in driving the transient BALs. For this, we constructed a control sample of radio quiet BALQSOs that have multiple epoch spectra in SDSS DR10. For each  QSO with transient \civ\ absorption 
in the radio detected sample, we constructed a control sample of 10 radio quiet BALQSOs matching closely in absolute i-band magnitude  and redshift. We first generated a sample of 20-25 radio quiet BALQSOs which are  closest in redshift and absolute i-band magnitude for each source in the radio detected transient \civ\ sample.  From this sample, we then selected  10 sources for each transient \civ\ BALQSO  which has multiple epoch spectra in SDSS DR10.  The typical dispersion in the $\Delta$z and  $\Delta$i magnitude values of the control sample is 0.02 and 0.2 respectively except for the case of J0959+6334 where it is 0.08 and 0.3. Each of these 60 sources in the control sample were first visually inspected for the presence of transient \civ\ BAL components. We then, followed the  procedure described in Section 3 to identify the transient \civ\ BAL components\footnote{Fig.\ref{fr_control_W_civ} shows the absolute fractional variation of \civ\ equivalent widths in the individual absorption 
components of the control sample sources with z$_{em}$, maximum velocity of the component, average equivalent widths and rest-frame time 
lags. We also find that our control sample  BALQSOs exhibit all the known correlations between BAL variability and various absorption line parameters discussed in Section 5.2. }.      Among the 60 sources in the control sample, 7 sources (i.e. 12$\pm$4 \%) show clear transient \civ\ features. 
{  This is close to 12$\pm$5 \% we find for our full radio detected sample with out applying any monitoring timescale cuts. Next, we consider only sources that are monitored over a period more than 800 days in the QSO's rest frame in both our's and the control sample. We find the transient detection rate of 23$\pm$9 \% and 18$\pm$7 \% for our and control sample respectively. This is also consistent within errors. The two rates are consistent with one another and roughly 30\% smaller than the detection rate we found for QSOs with log(L$_{radio}$) $>$ 32.9. However, the errors are too large to clearly favor radio brightness being an important driver of transient BAL flows. 

As ejection velocity is also an important quantity in the problem, we need to make sure the distribution of ejection velocity in the control sample is not very different from our radio detected sample.
 } 
Table~\ref{tab_emlist} gives a list of all the transient \civ\ BALQSOs identified in this study including those in the control sample.  The ejection velocity distributions of the radio detected and control sample BALQSOs are found to be similar. KS test between the two samples gives a D value of 0.13 and probability of 0.28. KS test between the transient 
\civ\ in the radio detected sample and the control sample gives the D and probability values of 0.3 and 0.06 respectively. The similarity of ejection velocity distributions in the above three samples ensures that  the 
transient \civ\ BAL phenomena in the control sample sources are not enhanced by the bias in the distribution of ejection velocity.
The lack of very strong correlation between the transient \civ\ BAL and the radio properties seem to suggest that there is no observable dependence between radio jets  and transient \civ\ absorbing flows in QSOs.        
 
\begin{figure*}
 \centering
\psfig{figure=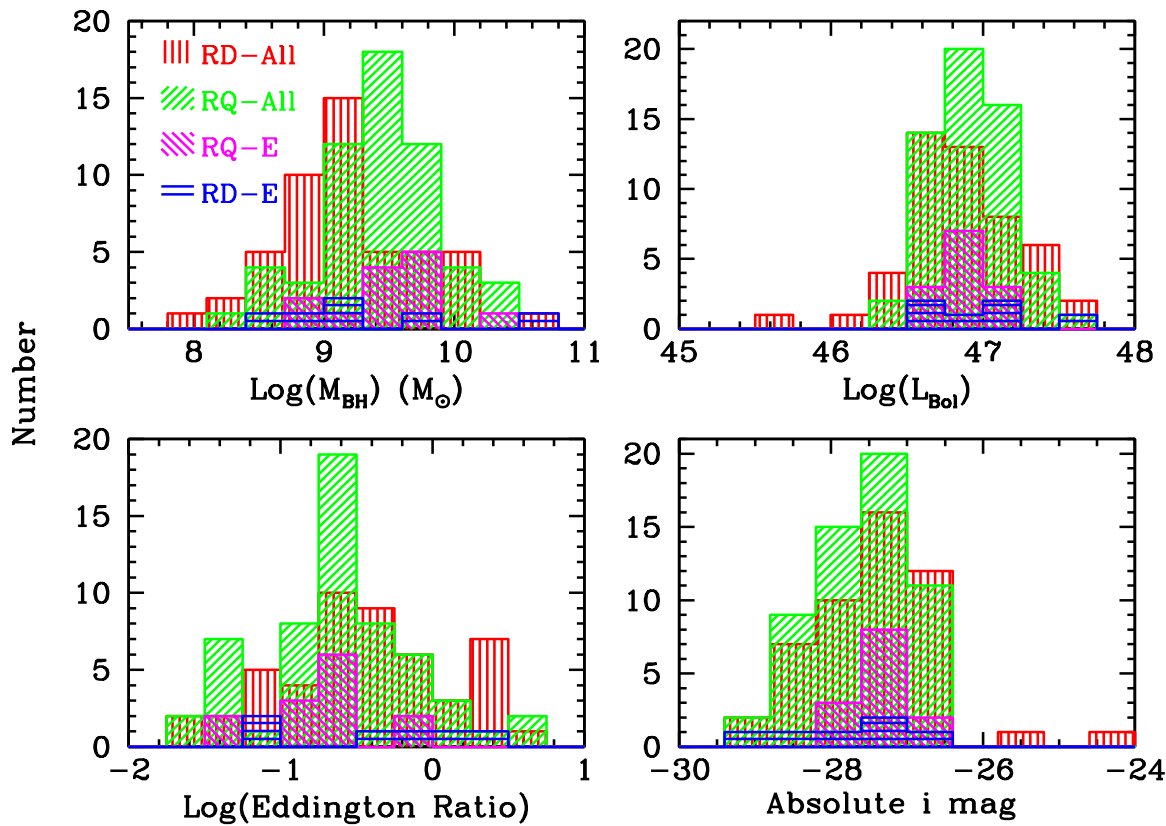,width=0.9\linewidth,height=0.65\linewidth,angle=270}
\caption{Histogram distribution of the quasar properties, Eddington ratio, mass of the blackhole, bolometric luminosity and absolute i-band magnitude for various samples in the study. {\it Red histogram with vertical lines }: radio detected sample, {\it green histogram with 45$^{\circ}$ slanted lines} : radio quiet control sample, {\it blue histogram with horizontal lines} : radio detected transient sample, {\it magenta histogram with -45$^{\circ}$ slanted lines} : radio quiet transient sample.     }

\label{hist_qsoprop}
\end{figure*} 

Fig.~\ref{hist_vel} gives a more clear picture of the trend with the ejection velocity.  In the figure, the top panel represents the histogram distribution of all the BAL components in the radio detected sample (blue/with slanted lines) and radio quiet control sample (red/with vertical lines). The middle panel presents the distribution of the transient \civ\ components and the lower panel represents the fraction of transient \civ\ BALs.
 It is clear that the fraction is high at higher velocities for both the radio detected and radio quiet control sample. We also note that the ratio of ejection velocity of the transient component to the maximum ejection velocity of the BALs peaks at unity. The 
transient \civ\ components are more likely to be with higher ejection velocities.  We also checked if any preferential time sampling of sources is the reason for the BAL transience in some of the sources in the sample.  Two-sided KS test for  the transient and  parent sample rest-frame time lag distributions result in  D and probability values of 0.18 and 0.17 respectively. This implies that the time-sampling of the sources with transient BALs is nothing special compared  to the other BALQSOs in the   sample.



\subsection{BAL transience and various derived QSO parameters}

To investigate further whether  basic properties of QSOs showing transient \civ\ BAL (both in our parent and control
sample) is different from the rest of the BALQSOs, we matched  the sources  in this study with those in the  SDSS DR7 QSO 
properties catalogue of 
\citet{shen11} to explore the correlation of transient \civ\ BALs with different inferred QSO properties.  
This catalogue covers all the sources in this study except for one non-transient source  each  in the 
radio detected and the control sample. Fig.~\ref{hist_qsoprop} shows the histogram distributions of the QSO 
properties: Eddington ratio, blackhole mass, bolometric luminosity and absolute i-band magnitude for the four 
samples in this study.  The red (with vertical lines), green (with 45$^{\circ}$ slanted lines), blue (with horizontal lines) and magenta (with -45$^{\circ}$ slanted lines) histograms represent the radio detected, radio detected transient, 
radio quiet, radio quiet transient sample. There is no evidence for any correlation between the transient \civ\
BALs with any of the above mentioned QSO properties. Results of KS tests also reveal that different 
samples in this study belong to the same `general' QSO population (see Table~\ref{tab_kstest}).

\subsection{Correlation with IR properties}
	We measure the far-infrared fluxes of our sources from WISE (Wide-field Infrared Survey Explorer) catalogue. 
	 WISE operates in four wavelength bands W1, W2, W3 and W4 which are centered at 3.4, 4.6, 12 and 22 microns respectively.  
We use  fluxes in these bands to  probe the role of dust in the transient outflows. Recently, \citet{zhang14} explored the role of dust in driving the outflows in a sample of 2099 BALQSOs. They find that the outflow strength and velocity is strongly correlated with the near infrared continuum slope which is a good indicator of dust emission. They proposed a  dusty outflow scenario  to explain the observed correlations between the near-infrared continuum slope and BAL strength. In this scenario, dusty clouds emerging from the outer regions of the accretion disk or innermost regions of the torus, are uplifted above the disk and exposed to the central engine. While the low density part of this cloud is highly ionized and is responsible for blue-shifted absorption lines, the high density part holds the dust and radiates in the NIR.
\begin{figure}
 \centering
\psfig{figure=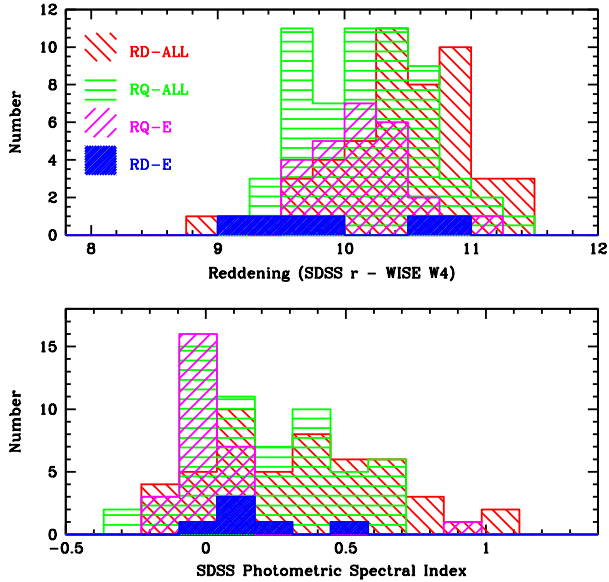,width=1.0\linewidth,height=1.0\linewidth,angle=0}
\caption{ Distribution of two parameters characterising reddening of the BALQSOs. {\it upper panel}: Reddening characterised by SDSS r - WISE W4 color.
{\it lower panel}: Reddening characterised by the SDSS photometric spectral index.  The four samples in this study are radio detected sample (red with -45$^{\circ}$ slanted lines), radio quiet control sample (green with horizontal lines), radio quiet transient \civ\ sample (magenta with 45$^{\circ}$ slanted lines)   and the radio detected transient \civ\ sample (blue filled).  }

\label{hist_reddening}
\end{figure}
 
   Several earlier studies in the literature have made use of a selection criteria in the  SDSS-WISE colors to pick dust obscured quasars.       
 Hot dust emission dominates the rest-frame SED of a QSO $>$ 3 $\mu m$.   Our sample has a minimum and maximum emission redshift of 1.67 and  3.63 with a median at 1.91.  WISE W4 magnitude clearly traces the dust emission for the redshift range of our sample.  In this study, we use the SDSS $r$ - WISE $W4$ color as a proxy for the strength of dust emission. Photometric spectral index values given in the SDSS catalogue can also be used as a measure of the reddening parameter.  Fig.~\ref{hist_reddening} gives the histogram distributions of the four samples in this study, namely radio detected sample (red with -45$^{\circ}$ slanted lines), radio quiet control sample (green with horizontal lines), radio quiet transient \civ\ sample (magenta with 45$^{\circ}$ slanted lines)   and the radio detected transient \civ\ sample (blue filled). From Fig.~\ref{hist_reddening}, there is no clear evidence for the role of dust  in driving the transient \civ\ BALs. Two sided KS test (see Table~\ref{tab_kstest}) for the radio detected and transient \civ\ BAL sample also does not point to any clear connection between the presence of dust and transient BAL phenomenon. Here again, we caution that our sample of radio detected BAL quasars is much smaller compared to the \citet{zhang14} sample.

\section{Discussion \& Summary}
 Recent studies have revealed that the radio properties of BAL  and non-BALQSOs are more alike than different. \citet{rawlings04} have found that whenever powerful jets are triggered, there is a dramatic increase in the efficiency of the feedback.  It is quite possible that  the onset of radio jets can have a feedback on to the accretion processes, which may trigger or quench the outflows.  The transient
\civ\ BAL systems in quasars can be used to determine the exact physical conditions prevailing at the time of outflow being ejected or quenched.

We have presented a detailed analysis of BAL variability for a sample of radio detected BALQSOs in the SDSS DR10 spectroscopic survey, with an emphasis on the transient \civ\ BALs. 
In the overall sample of radio detected sources, the fraction of transient \civ\ BAL sources (12\%)  is  higher than what is found (3.3\%) by \citet{filiz12} for disappearing \civ\ BALs. This may be due to the the  higher rest-frame observation timescales for our sample (6.5 yr) as compared to \citet{filiz12} sample (3.9 yr). This fraction is large (25\%) among the higher radio flux density sources in our sample. 
{ However, we notice that the occurrence of transient BALs in \citet{welling14} is not as high as this value. We show that the difference is mainly due to differences in the monitoring period of the two sample. When sub-samples are constructed with monitoring period larger than 800 days in the QSO rest-frame, we do  find higher detection rates even in the sample of \citet{welling14}. } 
{ Using a control sample of radio quiet BALQSOs matched to the QSOs showing transient \civ\ BALs in radio detected QSOs, we find 12\% cases showing transient \civ\ BAL troughs. This is consistent with the overall rate  found for radio detected sources.  Our control sample also hints on the occurrence of transient BALs being slightly higher in the case of radio bright QSOs. However, this difference is not overwhelmingly high and cannot be demonstrated at high statistical significance with the present data.}

Radio detected BALQSO sample in this study also follows various  BAL variability trends already known for \civ\ and \mgii\ BALs. Notable is the trend with the outflow velocity.  The transient \civ\ BALs most often occur at higher outflow velocities, typically greater than 10000 \kms. We also find that the transient BAL components are more frequently detected for rest-frame timescales of 800 days. This could either mean a characteristic timescale over which the ionization changes or the cloud crossing time across our line of sight. There is no correlation found between the transient BALs and optical continuum parameters. So, ionization changes in the outflowing gas due to the changing continuum is most unlikely   driver for the transient phenomena.  However, it is possible that  the optical continuum variation inferred with CRTS lightcurves obtained without a specific filter may not be an ideal tracer of the ionizing continuum.   We also do not find any correlations between the  occurrence of transient \civ\ BALs  and various other QSO properties like Eddington ratio, black hole mass and luminosity.

{ We use the characteristic ejection velocities (i.e $>$ 10000 \kms) and timescale ($\sim$ 800 days) of the transient flow with a simple model proposed by \citet{hall11} to place average constraints on the location and transverse speed of \civ\ emerging BAL components. Following \citet{hall11} , we estimate the  diameter of the disc within which 90 per cent of the 2700 $\AA$ continuum is emitted to be D$_{2700}$ $\sim$ 2 $\times$ 10$^{16}$ cm for the typical black hole of mass 10$^9$M$\odot$ for our sample. This, together with the characteristic BAL transience time scale of 800 rest frame days results in a  transverse velocity of 3100 \kms. We assume a typical line of sight (LOS) velocity of 1$\times$10$^4$ \kms for our calculations. Section 4.3 and 4.4 of \citet{hall11} give a model for constraining the BAL structure location.  The exact estimate of the location of the flow  requires the knowledge of the angle of LOS $\lambda$ and the angle of the velocity vector $\theta$ above the accretion disc. As we are interested in the average properties, we computed the location of the BAL outflow for each combination of $\lambda$ and $\theta$ and then averaged it. Following the procedure of \citet{hall11}, we obtain an estimate of BAL launching radius to be 0.5$\pm$0.001 pc and the distance d$_{BAL}$ from the black hole to be 2.2$\pm$0.1 pc. This is just outside the broad line region ($\sim$ 1 pc) for a typical quasar of log(L$_{bol}$) = 46.8 ergs s$^{-1}$.  Thus, the flow has not yet reached ISM to provide a strong feedback on the star formation.}

In the absence of clear evidences for any correlations between the occurrence of transient \civ\ BALs and various parameters explored here, it is probable that transient BAL phenomenon  is just an extension of normal BAL variability. {  Similar  optical continuum variability  and BAL transience properties of radio detected and radio quiet BAL QSO samples might suggest that radio detected BAL QSOs are not necessarily always viewed down the jet axis. Our results do not support the case of jet-cloud interactions triggering bulk motions in the BAL outflow that can be seen in the form of transient BAL components.}
We also realize that  the number transient BALs in our sample is too small to extract any possible weak dependences. A similar study with a  larger sample size is needed to completely understand the transient BAL phenomenon. This study increases the sample size of transient BALQSOs by 18. Future spectroscopic monitoring of these sources will allow us to look for further dynamical evolution in these BALs. Multi wavelength observations of these sources will help to understand the role of changing 'sheilding gas' in driving transient BAL phenomenon.   
Ongoing and future spectroscopic surveys like eBOSS (extended Baryon Oscillation Spectroscopic Survey), LAMOST (Large Sky Area Multi-Object Fibre Spectroscopic Telescope) look promising for identifying new transient BAL sources and understanding the physical mechanism driving the transient phenomenon.

\section*{acknowledgements}
{We thank the anonymous referee for a thorough review and helpful comments which led to an improved manuscript. Funding for SDSS-III has been provided by the Alfred P. Sloan Foundation, the Participating Institutions, the National Science Foundation, and the U.S. Department of Energy Office of Science. The SDSS-III web site is http://www.sdss3.org/.}

{SDSS-III is managed by the Astrophysical Research Consortium for the Participating Institutions of the SDSS-III Collaboration including the University of Arizona, the Brazilian Participation Group, Brookhaven National Laboratory, Carnegie Mellon University, University of Florida, the French Participation Group, the German Participation Group, Harvard University, the Instituto de Astrofisica de Canarias, the Michigan State/Notre Dame/JINA Participation Group, Johns Hopkins University, Lawrence Berkeley National Laboratory, Max Planck Institute for Astrophysics, Max Planck Institute for Extraterrestrial Physics, New Mexico State University, New York University, Ohio State University, Pennsylvania State University, University of Portsmouth, Princeton University, the Spanish Participation Group, University of Tokyo, University of Utah, Vanderbilt University, University of Virginia, University of Washington, and Yale University. }
 
\def\aj{AJ}%
\def\actaa{Acta Astron.}%
\def\araa{ARA\&A}%
\def\apj{ApJ}%
\def\apjl{ApJ}%
\def\apjs{ApJS}%
\def\ao{Appl.~Opt.}%
\def\apss{Ap\&SS}%
\def\aap{A\&A}%
\def\aapr{A\&A~Rev.}%
\def\aaps{A\&AS}%
\def\azh{AZh}%
\def\baas{BAAS}%
\def\bac{Bull. astr. Inst. Czechosl.}%
\def\caa{Chinese Astron. Astrophys.}%
\def\cjaa{Chinese J. Astron. Astrophys.}%
\def\icarus{Icarus}%
\def\jcap{J. Cosmology Astropart. Phys.}%
\def\jrasc{JRASC}%
\def\mnras{MNRAS}%
\def\memras{MmRAS}%
\def\na{New A}%
\def\nar{New A Rev.}%
\def\pasa{PASA}%
\def\pra{Phys.~Rev.~A}%
\def\prb{Phys.~Rev.~B}%
\def\prc{Phys.~Rev.~C}%
\def\prd{Phys.~Rev.~D}%
\def\pre{Phys.~Rev.~E}%
\def\prl{Phys.~Rev.~Lett.}%
\def\pasp{PASP}%
\def\pasj{PASJ}%
\def\qjras{QJRAS}
\def\rmxaa{Rev. Mexicana Astron. Astrofis.}%
\def\skytel{S\&T}%
\def\solphys{Sol.~Phys.}%
\def\sovast{Soviet~Ast.}%
\def\ssr{Space~Sci.~Rev.}%
\def\zap{ZAp}%
\def\nat{Nature}%
\def\iaucirc{IAU~Circ.}%
\def\aplett{Astrophys.~Lett.}%
\def\apspr{Astrophys.~Space~Phys.~Res.}%
\def\bain{Bull.~Astron.~Inst.~Netherlands}%
\def\fcp{Fund.~Cosmic~Phys.}%
\def\gca{Geochim.~Cosmochim.~Acta}%
\def\grl{Geophys.~Res.~Lett.}%
\def\jcp{J.~Chem.~Phys.}%
\def\jgr{J.~Geophys.~Res.}%
\def\jqsrt{J.~Quant.~Spec.~Radiat.~Transf.}%
\def\memsai{Mem.~Soc.~Astron.~Italiana}%
\def\nphysa{Nucl.~Phys.~A}%
\def\physrep{Phys.~Rep.}%
\def\physscr{Phys.~Scr}%
\def\planss{Planet.~Space~Sci.}%
\def\procspie{Proc.~SPIE}%
\let\astap=\aap
\let\apjlett=\apjl
\let\apjsupp=\apjs
\let\applopt=\ao
\bibliographystyle{mn}
\bibliography{refe}

\appendix
\section{Control Sample : Emerging sample}
\begin{figure*}
 \centering
\psfig{figure=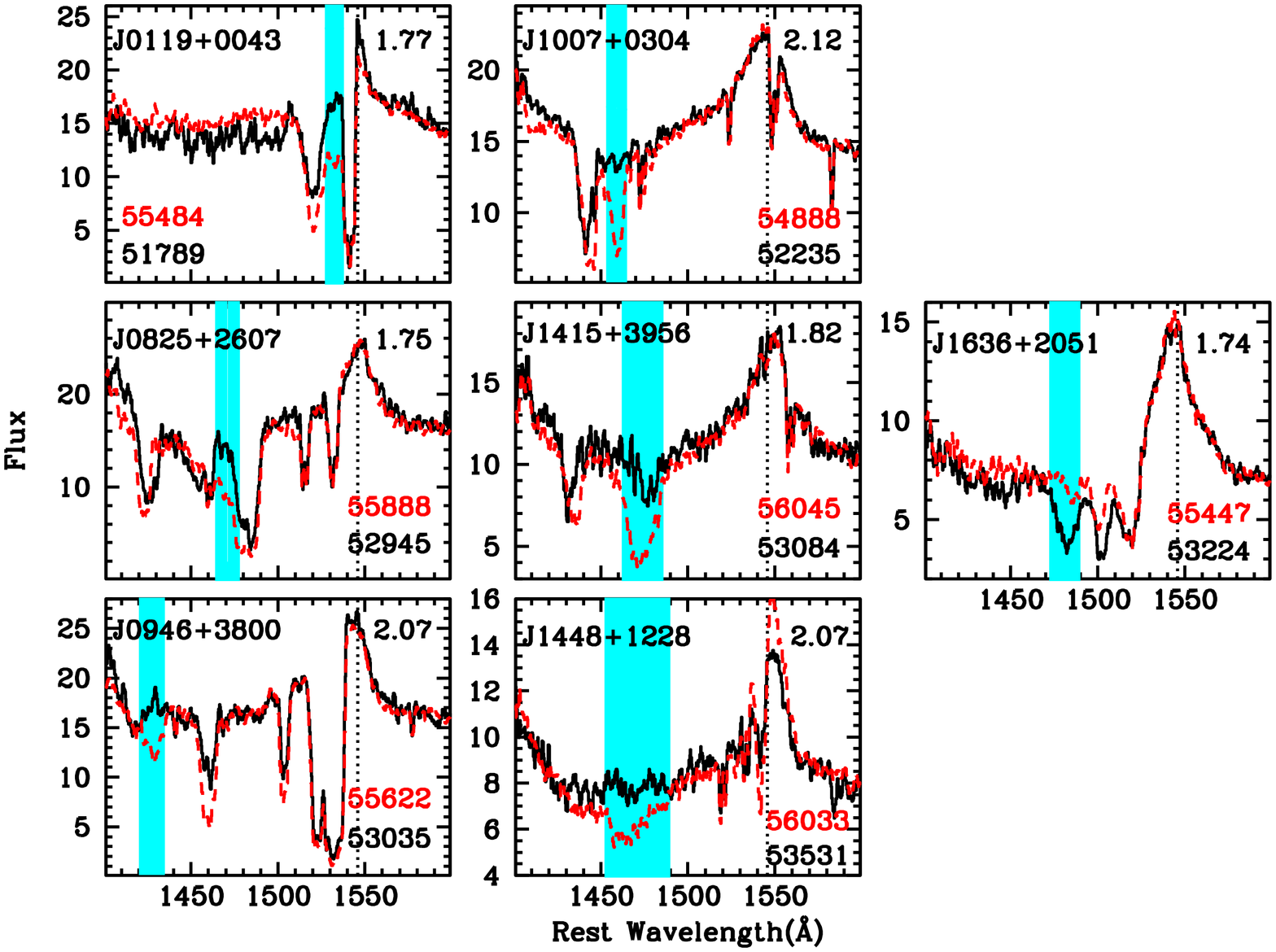,width=1.0\linewidth,height=0.8\linewidth,angle=270}
\caption{   Spectra of control sample sources showing BAL appearance/disappearance. Red dashed spectra correspond to later epoch measurements. Shaded region represents the transient components.  Relative velocity is measured from the \civ\ emission line. Redshift, epoch of observations  for each source are also marked.  }
\label{cs_emerge_plot}
\end{figure*}

\begin{figure*}
 \centering
\begin{tabular}{c c}
\psfig{figure=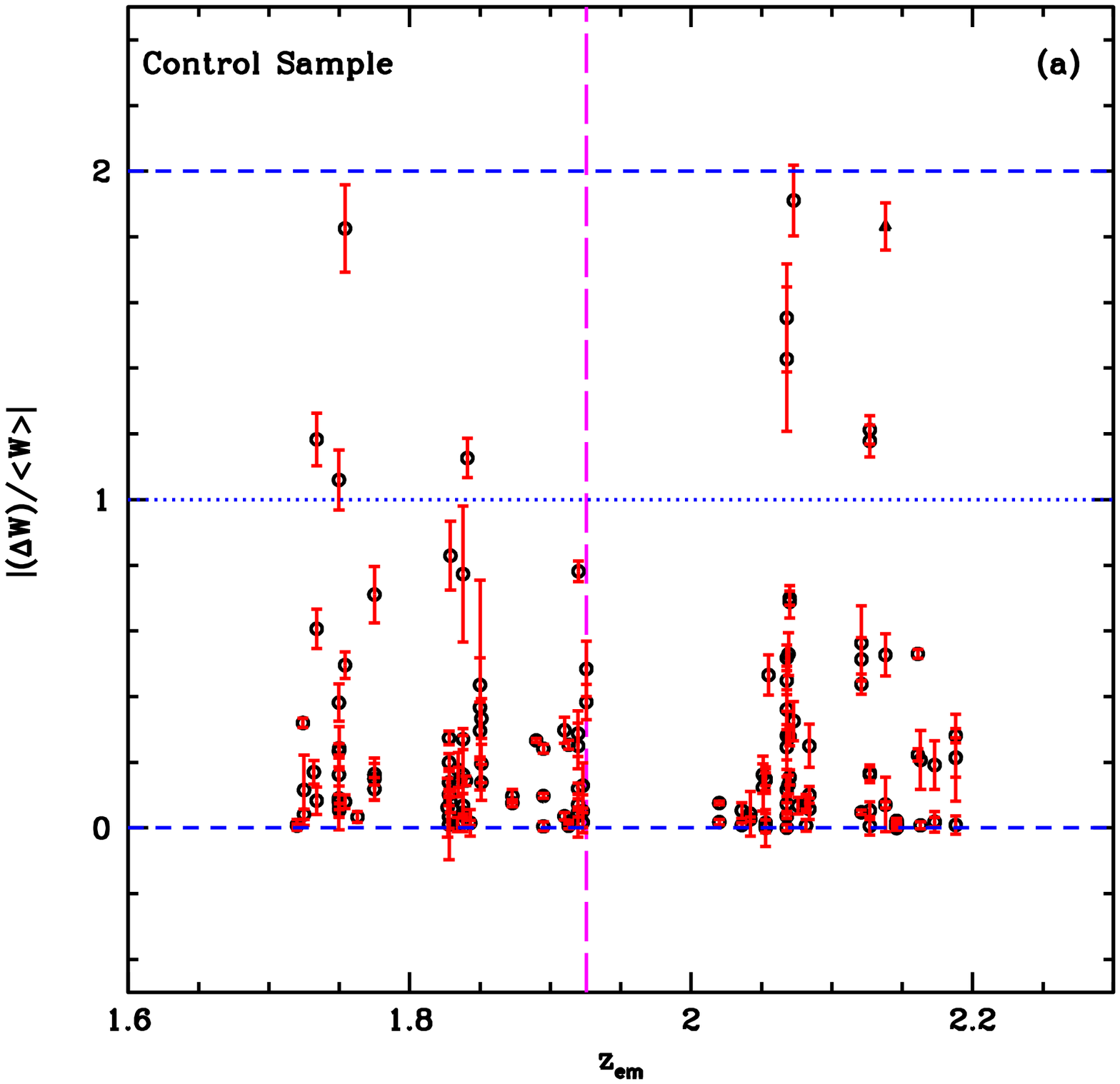,width=0.5\linewidth,height=0.35\linewidth,angle=0}&
\psfig{figure=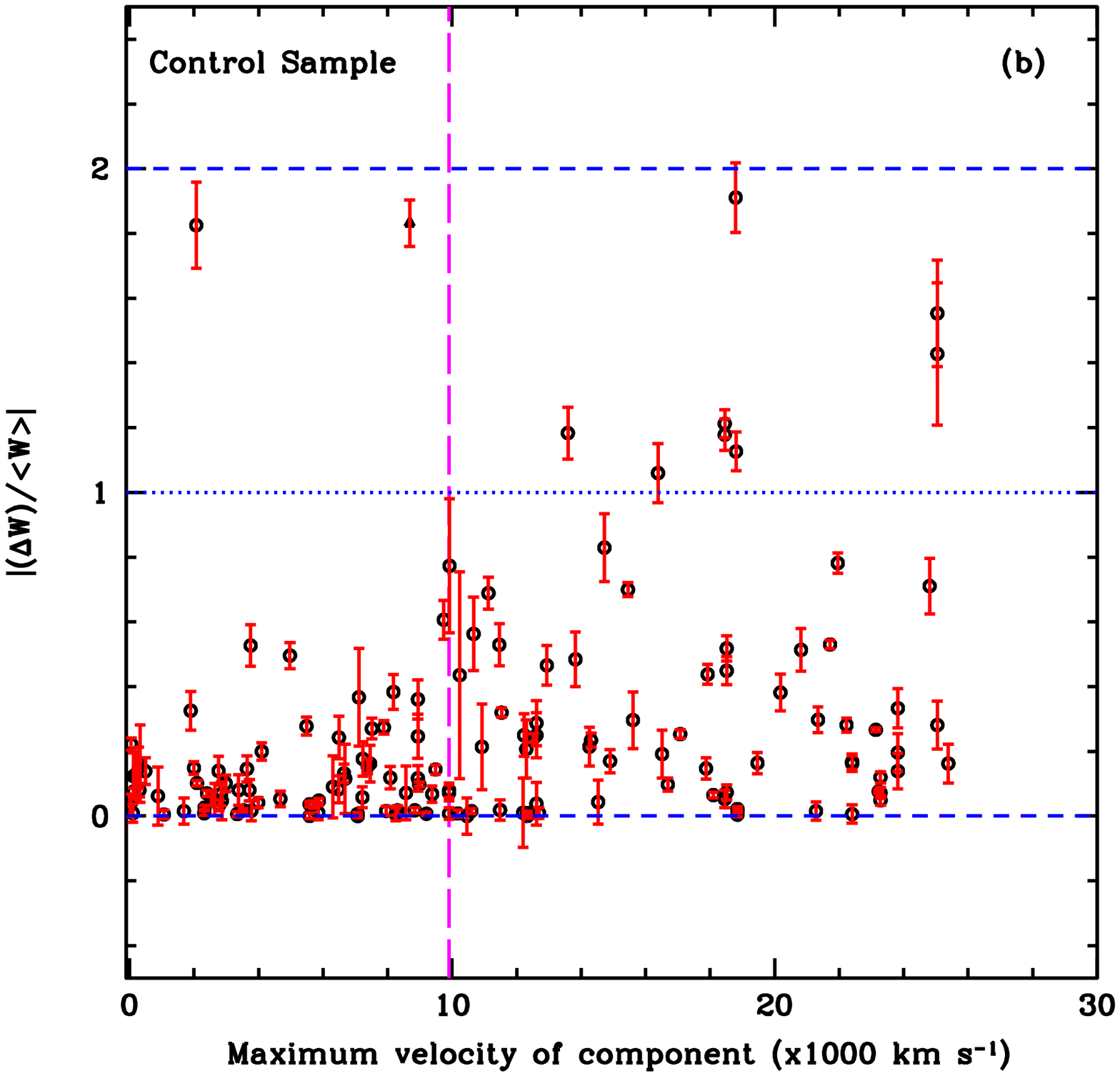,width=0.5\linewidth,height=0.35\linewidth,angle=0}\\
\psfig{figure=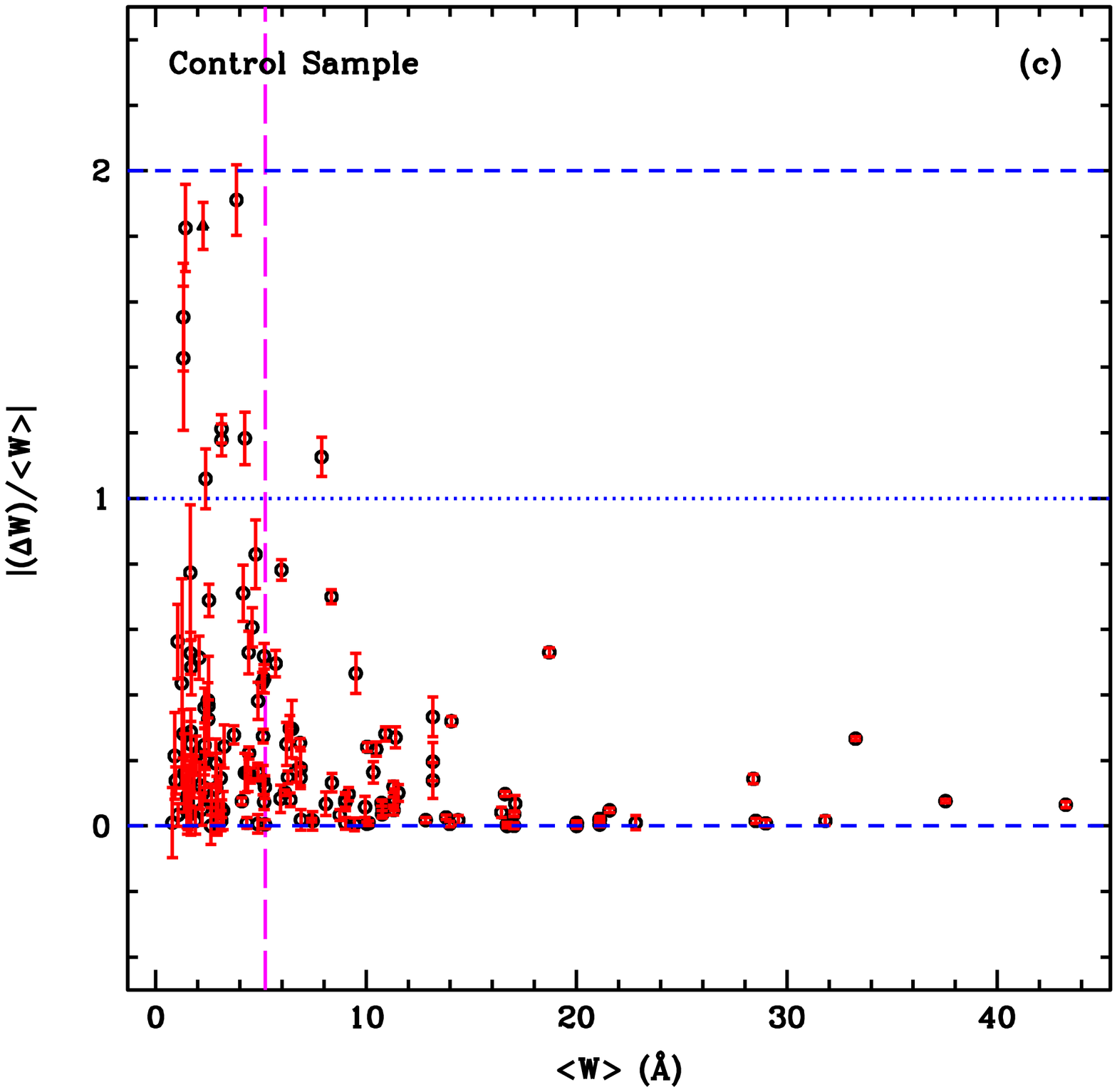,width=0.5\linewidth,height=0.35\linewidth,angle=0}&
\psfig{figure=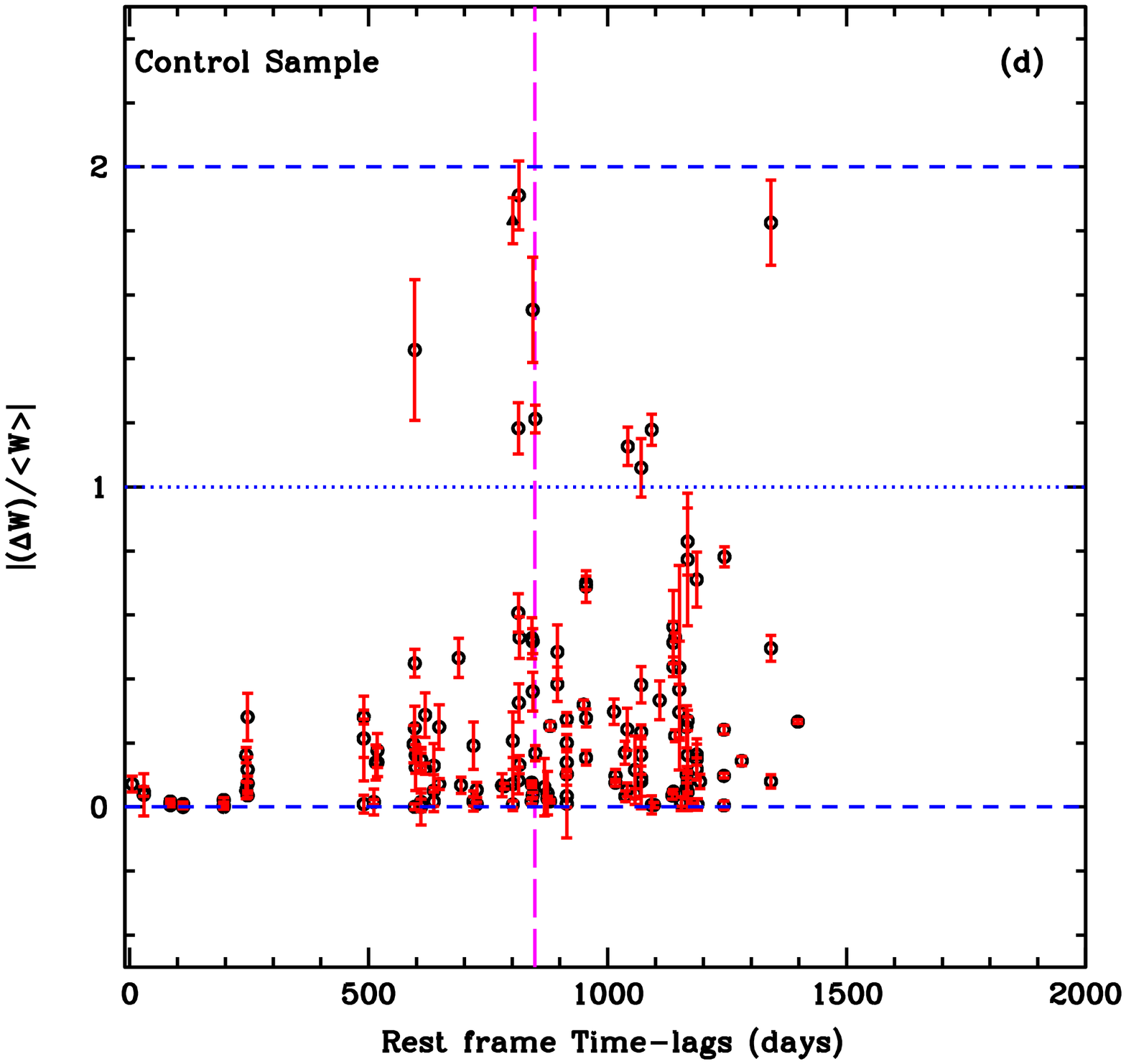,width=0.5\linewidth,height=0.35\linewidth,angle=0}\\

\end{tabular}
\caption{Absolute fractional variation of \civ\ equivalent widths of the control sample sources are plotted against emission redshift, maximum velocity of the component, fractional equivalent width variation, and rest-frame timelags. The upper blue horizontal dashed line represents a fractional variation of 2 which corresponds to emergence of BAL. The blue dotted horizontal line represents a fractional variation of 1. The median value of the distributions are marked by the vertical long dashed magenta line in each plot. }
\label{fr_control_W_civ}
\end{figure*}



\end{document}